\documentclass[12pt,manuscript]{aastex}
\usepackage{soul}
\slugcomment{}

\shorttitle{}
\shortauthors{}

\begin{document}

\title{ON THE EVOLUTION OF ANOMALOUS X-RAY PULSARS AND SOFT GAMMA RAY REPEATERS WITH FALLBACK DISKS}
%\author { \"{U}. ERTAN\altaffilmark{1}, M. H. ERKUT\altaffilmark{2}}
\author{\"{U}. Ertan\altaffilmark{1}, K.Y. Ek\c{s}i\altaffilmark{2}, M.H. Erkut\altaffilmark{3}  \&  M.A. Alpar\altaffilmark{1} }
%\altaffilmark{1},
\affil{\altaffilmark{1}Sabanc\i\ University, 34956, Orhanl\i\, Tuzla, \.Istanbul, Turkey}
\affil{\altaffilmark{2}\.Istanbul Technical University, Faculty of Science and Letters, Physics Engineering Department, FE2, Maslak, 34469, \.Istanbul, Turkey}
\affil{\altaffilmark{3}Department of Mathematics and Computer Science,
\.{I}stanbul K\"{u}lt\"{u}r University, Bak\i rk\"{o}y 34156, \.{I}stanbul, Turkey}

%\email{ertan@sabanciuniv.edu, hakane@sabanciuniv.edu, yavuz@sabanciuniv.edu, %alpar@sabanciuniv.edu}

%\altaffiltext{1} { \.Istanbul Technical University %(\.{I}T\"{U}), \.Istanbul, Turkey.}

\def\la{\raise.5ex\hbox{$<$}\kern-.8em\lower 1mm\hbox{$\sim$}}
\def\ga{\raise.5ex\hbox{$>$}\kern-.8em\lower 1mm\hbox{$\sim$}}
\def\be{\begin{equation}}
\def\ee{\end{equation}}
\def\ba{\begin{eqnarray}}
\def\ea{\end{eqnarray}}
\def\m{\mathrm}
\def\d{\partial}
\def\R{\right}
\def\L{\left}
\def\a{\alpha}
\def\Mdot*{\dot{M}_*}
\def\Mdotin{\dot{M}_{\mathrm{in}}}
\def\Mdot{\dot{M}}
\def\Pdot{\dot{P}}
\def\Msun{M_{\odot}}
\def\Lin{L_{\mathrm{in}}}
\def\Rin{R_{\mathrm{in}}}
\def\rin{r_{\mathrm{in}}}
\def\rout{r_{\mathrm{out}}}
\def\rlc{r_{\mathrm{lc}}}
\def\Rout{R_{\mathrm{out}}}
\def\Ldisk{L_{\mathrm{disk}}}
\def\Lx{L_{\mathrm{x}}}
\def\cs{c_{\mathrm{s}}}
\def\dEb{\delta E_{\mathrm{burst}}}
\def\dEx{\delta E_{\mathrm{x}}}
\def\deltam{\delta{\mathrm{m}}}
\def\Bb{\beta_{\mathrm{b}}}
\def\Be{\beta_{\mathrm{e}}}
\def\Rc{\R_{\mathrm{c}}}
\def\rA{r_{\mathrm{A}}}
\def\rp{r_{\mathrm{p}}}
\def\Tp{T_{\mathrm{p}}}
\def\dMin{\delta M_{\mathrm{in}}}
\def\dM*{\delta M_*}
\def\Teff{T_{\mathrm{eff}}}
\def\Tirr{T_{\mathrm{irr}}}
\def\Firr{F_{\mathrm{irr}}}
\def\Tcrit{T_{\mathrm{crit}}}
\def\P0min{P_{0,{\mathrm{min}}}}
\def\Av{A_{\mathrm{V}}}
\def\ah{\alpha_{\mathrm{hot}}}
\def\ac{\alpha_{\mathrm{cold}}}
\def\p{\propto}
\def\m{\mathrm}
\def\fast{\omega_{\ast}}

\begin{abstract}

We show that the period
clustering of anomalous X-ray pulsars (AXPs) and soft gamma-ray repeaters (SGRs), their X-ray luminosities,  ages and statistics can be explained with fallback disks with large initial specific angular momentum. 
The disk evolution models are developed by comparison to self-similar analytical models. The initial disk mass and angular momentum set the viscous timescale. An efficient torque, with $(1 - \fast^2)$ dependence on the fastness parameter $\fast$ leads to period clustering in the observed AXP-SGR period range under a wide range of initial conditions.  The timescale $t_0$ for the early evolution of the fallback disk, and the final stages of fallback disk evolution, when the disk becomes passive, are the crucial determinants of the evolution.    
The disk becomes passive at temperatures around
100 K, which provides a natural cutoff for the X-ray luminosity
and defines the end of evolution in the observable AXP and SGR phase. This low value for the minimum temperature for active disk turbulence indicates that the fallback disks are active up to a large radius, $\ga ~10^{12}$ cm. 
We find that transient AXPs and SGRs  are likely to be older than their
persistent cousins. A fallback disk with mass transfer rates corresponding to the low quiescent X-ray
luminosities of the transient sources in early evolutionary phases would have
a relatively lower initial mass, such that the mass-flow rate in the disk is not sufficient for the inner disk 
to penetrate into the light cylinder of the young neutron star,  making mass accretion onto the neutron star impossible. The transient AXP phase therefore must start later. The model results imply that the transient AXP/SGRs, although older, are likely to be similar in number to persistent sources. This is because the X-ray luminosities of AXPs and SGRs are found to decrease faster at the end of their evolution, and the X-ray luminosities of transient AXP and SGRs in quiescence lie  in the luminosity range of X-ray cutoff phase. Taking the range of quiescent X-ray luminosities of transient AXPs and SGRs~ $\sim 10^{33} - 10^{34}$ erg s$^{-1}$, our simulations imply that the duration of  the cutoff phase is comparable to the lifetime in the persistent phase for a large range of initial conditions.   
   
\end{abstract}


\section{INTRODUCTION}

Anomalous X-ray pulsars (AXPs) and soft gamma-ray repeaters (SGRs) 
\citep[see][for a review]{WT05,mereg08} are rapidly spinning down young neutron stars
with persistent X-ray luminosities $\Lx \sim 10^{34} - 10^{36}$ erg s$^{-1}$ 
in excess of their spin-down power.  
The spin periods of this small population with $\sim 15$ members is 
clustered between $P=2-12$ s; placing them in the top right corner 
of the $P-\dot{P}$-diagram, populated by only a few radio pulsars.

The recognition of AXPs and SGRs among the new classes of young objects, along with other young neutron star families which do not manifest themselves as radio pulsars, like the dim isolated neutron stars (DINs) and the compact central objects (CCOs), suggests alternative evolutionary paths for young neutron stars. The magnetar model \citep{DT92,TD95} attributes SGR and AXP bursts and other properties to  magnetic field stronger than quantum critical field, 
$B_{\rm c} = 4.4 \times 10^{13}$ G. The presence of high magnetic field radio pulsars with super-critical inferred fields challenges this being the only criterion. An alternative \citep{alpar01,CHN} {\emph or} complementary \citep{EkA03,ErA03} 
view suggests the presence of remnant disks around 
young neutron stars left over from the supernova explosion as a necessary ingredient 
causing the diversity.  The initial mass {\em and} angular momentum of the disk present a new set of parameters, in addition to the initial period and dipole
 magnetic moment of the neutron star, that 
 affect evolutionary paths and observational classes of neutron stars.

AXPs and SGRs show bursts with luminosities well above the Eddington limit.  
This cannot be explained by gas accretion onto the neutron star.
The magnetar model explains these bursts as instabilities in the crust due to 
strong magnetic stresses and
reconfiguration of the magnetosphere, while it accounts for the observed spin-down torques of AXP and SGRs by  magnetic dipole radiation.  
In the magnetar model it is hard to explain the period 
clustering of AXPs and SGRs within the framework that the persistent X-ray luminosity 
is a consequence of the magnetic field decay \citep{PM02}.

The detection of a debris disk around the AXP 4U 0142+61 \citep{WCK06} could be the clue for what makes AXPs and SGRs different 
from radio pulsars in their persistent phases \citep{EEEA07}.
The presence of fallback disks can address the period clustering of AXPs and SGRs 
as a result of the long-term interaction of the disk with the magnetic dipole field of the neutron star, converging to "equilibrium" periods. 
While the period clustering generated by the earlier models is in agreement with observations, 
there were difficulties in explaining the spin-periods together with X-ray luminosities \citep{EkA03}.
 
In this paper, we have identified three crucial aspects of fallback disk evolution: (1) The magnetic torque on a "mildly" fast neutron star in the regime of accretion with spin-down, has the dependence ~$N\propto (1 - \fast^2)$  on the fastness parameter $\fast$ (see also \citet{EE08}). (2) The disk timescale ~$t_0$~ that determines the initial evolution of the system is not arbitrary, but instead set by the initial mass and angular momentum of the disk. (3) The outer regions of the disk become passive at temperatures ~$T < \Tp \sim 100$ K. This results in the mass-inflow rate $\Mdot$ of the disk being turned off rather sharply at ages $\ga$ a few $10^4 - 10^5$ y. In \S\ 2, we present the disk equations and self-similar models for fallback disk evolution. The timescale $t_0$, which is particularly important for early disk evolution is related to the initial disk mass and angular momentum. We also briefly discuss the assumptions and results of earlier analytical work on the long-term evolution of fallback disks.
Comparison of analytical and numerical models is used to set the stage for the numerical calculations detailed in \S\ 3. This section also presents the torque model employed in our calculations and the justifications for this choice of torque on the basis of work on transient AXPs. The late evolution of the fallback disk is also addressed in \S\ 3.   
Fallback disk models used in evolutionary calculations so far  \citep{MPH01,EkA03} employ a 
gradual decrease of the mass-flow rate, $\dot{M}$, at late ages, leading to a population of old observable 
sources with periods extending to $\sim 100$ s contrary to the cutoff period $P\cong 15$ s inferred from 
an analysis of period statistics \citep{PM02}. Physically, as a disk evolves to decreasing $\dot{M}$, 
it will cool starting from the outer disk. For a while irradiation from the neutron star 
may keep the disk sufficiently hot. As $\dot{M}$ decreases even irradiation will not sustain high temperatures. 
One expects that eventually the ionization fraction in disk matter will be so low that the magneto-rotational 
instability (MRI) (Balbus \& Hawley 1991) which is believed to provide the turbulent disk viscosity should fail to operate. 
The disk in late phases should make a transition to an effectively neutral, passive state starting from the outer disk and leading to 
a cutoff in $\dot{M}$ and luminosity. 
We discuss the results of model calculations in \S\ 4 and summarize our conclusions in \S\ 5.

\section{EVOLUTION OF VISCOUS THIN DISKS IN TERMS OF INITIAL MASS AND ANGULAR MOMENTUM}

The evolution of the surface mass density $\Sigma$ in a viscous thin disk \citep{pringle81}
is described by the diffusion equation
\begin{equation}
\frac{\partial \Sigma }{\partial t}=\frac{3}{r}\frac{\partial }{\partial r}%
\left[ r^{1/2}\frac{\partial }{\partial r}(\nu \Sigma r^{1/2})\right]
\label{diffusion_eq}
\end{equation}
\citep{lust52} where $\nu$ is the turbulent viscosity. 
For opacities of the form $\kappa =\kappa_{0}\rho ^{a}T^{b}$
one can solve the disk structure equations to obtain viscosity in the form \citep[see][]{CLG}
\begin{equation}
\nu =Cr^{p}\Sigma ^{q}  \label{eq2}
\end{equation}%
where $C$, $p$ and $q$ are constants given by 
\begin{eqnarray}
C &=&\alpha^{\frac{2}{6-2b+a}+1}\left( \frac{27\kappa _{0}}{32\sigma
_{\rm SB}}\right) ^{\frac{2}{6-2b+a}}\left( \frac{k_{\rm B}}{\bar{\mu}m_{\rm p}}\right) ^{%
\frac{2-a}{6-2b+a}+1}\left( GM_{\ast }\right) ^{\frac{a+1}{6-2b+a}-\frac{1}{2%
}},  \label{C} \\
p &=&-3\left( \frac{a+1}{6-2b+a}-\frac{1}{2}\right),  \label{p} \\
q &=&\frac{2\left( a+2\right) }{6-2b+a}.  \label{q}
\end{eqnarray}
Here $M_{\ast}$ is the mass of the neutron star, $\alpha$ is the viscosity parameter \citep{SS73}, $\bar{\mu}$ is the mean molecular weight,
$k_{\rm B}$ is Boltzmann constant, $m_{\rm p}$ is the mass of the proton, and $\sigma
_{SB}$ is the Stephan-Boltzmann constant.

In general $q \neq 0$ and $\nu$ depends on the surface mass density and the diffusion equation (\ref{diffusion_eq}) is nonlinear. 
Three self-similar solutions were
given by \citet{pringle74} corresponding to different boundary conditions. 
The solution of interest here is the one corresponding to a disk of
constant total angular momentum and decreasing mass. This analytical model is a useful guide 
for the evolutionary calculation for the neutron star's spin, since the total angular momentum 
transfer from the neutron star to the disk is negligible compared to the angular momentum of the disk.
The original solutions as given by \citet{pringle74} diverge at $t=0$.
As the diffusion equation is symmetric with respect to translations in time, we can shift the origin of time
by $t \rightarrow t+t_0$ for all occurrences of $t$ in the solutions.
With this modification the solution becomes
\begin{equation}
\frac{\Sigma }{\Sigma _{0}}=k\left( 1+\frac{t}{t_{0}}\right) ^{-\frac{5}{%
5q-2p+4}}\left( \frac{r}{R_{\mathrm{out}}\left( t\right) }\right) ^{-\frac{p%
}{q+1}}\left[ 1-\left( \frac{r}{R_{\mathrm{out}}\left( t\right) }\right) ^{2-%
\frac{p}{q+1}}\right] ^{1/q}
\label{solution}
\end{equation}
where 
\begin{equation}
R_{\mathrm{out}}\left( t\right)
=r_{0}\left( 1+\frac{t}{t_{0}}\right) ^{\frac{2}{5q-2p+4}}
\end{equation}
is the outer boundary of the disk and
\begin{equation}
k=\left( \frac{q}{\left( 4q-2p+4\right) \left(5q-2p+4\right) }\right)^{1/q}.
\end{equation}
In these equations $\Sigma_0$, $t_0$ and $\nu_0$
are constants introduced in the non-dimensionalization procedure 
and will be defined below.

Using the solution in equation (\ref{solution}),  the total angular momentum 
of the disk is
\begin{equation}
J_{\rm d}=\int r^{2}\Omega _{K}\Sigma \cdot 2\pi rdr=\gamma _{1}4\pi r_{0}^{2}\Sigma _{0}\sqrt{GMr_{0}}
\end{equation}%
a constant. Here
\begin{equation}
\gamma_1=k\frac{q+1}{4q-2p+4}\beta (\frac{q+1}{q},\frac{5q-2p+5}{4q-2p+4})
\end{equation}
is a constant where 
$\beta (k,l)=\Gamma (k)\Gamma (l)/\Gamma (k+l)$ 
is the beta function.
The total mass of the disk
\begin{equation}
M_{\rm d}=\int \Sigma \cdot 2\pi rdr =M_{0}\left( 1+\frac{t}{t_0}\right)^{-\frac{1}{5q-2p+4}}
\end{equation}
is a decreasing function where
\begin{equation}
M_0=4\pi r_0^2\Sigma_0\gamma_{2} 
\end{equation}
and $\gamma_2=kq/(4q+4-2p)$.
Accordingly, the mass flow rate at the inner disk is
\begin{equation}
\dot{M}_{\rm d}=\frac{(a-1)M_0}{t_0} \left( 1+\frac{t}{t_0}\right)^{-a}
\label{Mdot}
\end{equation}
where $a = 1+1/(5q-2p+4)$.
For a disk of given initial mass $M_0$ and angular momentum $J_0$ the other
scale factors can be found through the following sequence: 
\begin{eqnarray}
r_0 &=&\frac{( J_0/M_0)^2}{GM}( \frac{\gamma _{2}}{\gamma _{1}}) ^{2} \\
\Sigma_0 &=&\frac{M_0}{4\pi r_0^2\gamma_2} \\
\nu_0 &=& C r_0^p \Sigma_0^q \\
t_0 &=&\frac{4r_0^2}{3\nu_0} 
\label{scalings}
\end{eqnarray}

\begin{figure}[t]
\epsscale{1.0} 
\plottwo{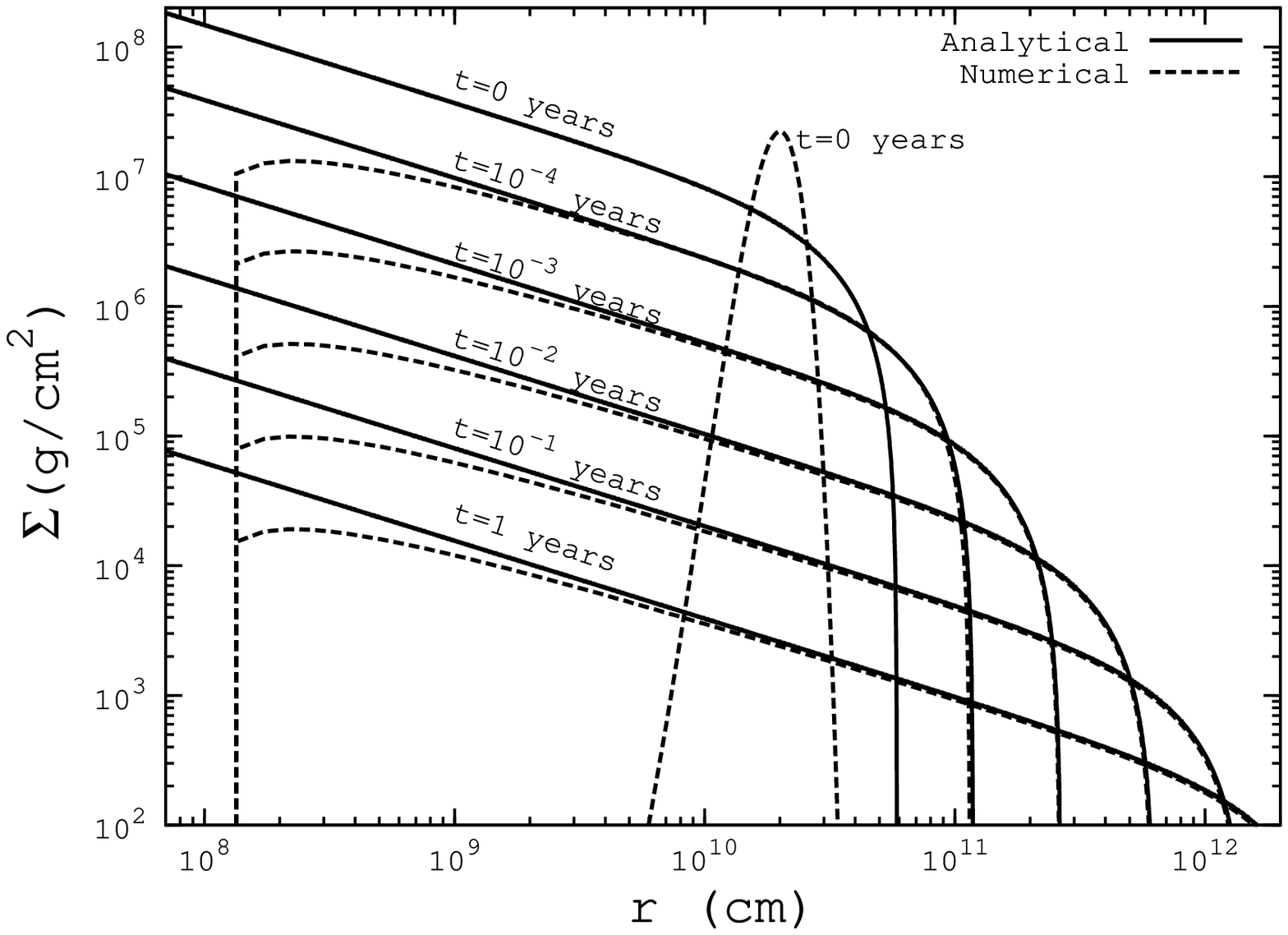}{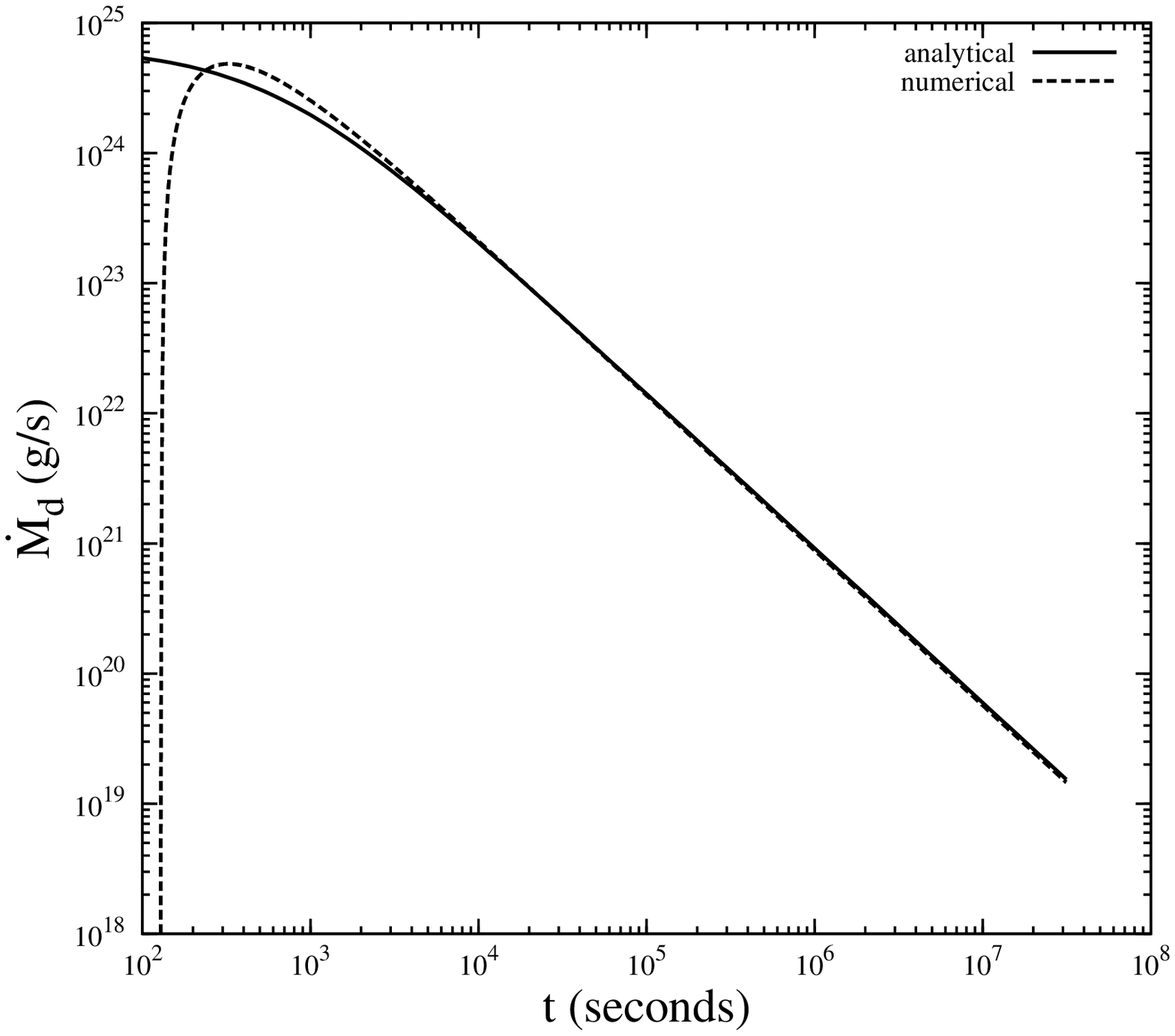}
\caption{Comparison of numerical solution of the diffusion equation with the analytical solution given in equation \ref{solution}. The initial surface mass density distribution is a gaussian $\Sigma(r,0)=\Sigma_{\max}\exp(-(r-r_{\max})^2/\sigma^2)$ where $r_{\max}=2\times 10^{10}$ cm, $\sigma=4 \times 10^9$ cm and $\Sigma_{\max}=2.23\times 10^7$ g cm$^{-2}$ is chosen such that the initial mass of the disk is $M_0=10^{-5}M_{\sun}$.
The corresponding initial angular momentum of the disk is $J_0=1.03\times 10^{47}$ g cm$^2$ s$^{-1}$. The initial mass and angular momentum of the disk yield the following non-dimensionalization scales: $\Sigma_0=7.35\times 10^8$  cm$^{-2}$, $r_0=5.86\times 10^{10}$ cm and $t_0=5.74\times 10^2$ y.
Left Panel: Evolution of the surface mass density $\Sigma$. 
Right Panel: Evolution of the mass flow rate $\dot{M}_d$ at the inner radius.}
\label{fig_1}
\end{figure}

\citet{CLG} showed that the self-similar solutions of  \citet{pringle74} corresponds
to the intermediate asymptotic \citep{barenblatt96} behavior of a disk with the no-viscous-torque
boundary condition at the inner radius ensured by $\Sigma (R_{\rm in},t)=0$.
In Figure~\ref{fig_1}, we compare the analytical solution given in equation~(\ref{solution})
with the numerical 
solution of the diffusion equation with a Gaussian initial distribution discretized on 2000 grids over a numerical domain extending from $10^8$ cm to $10^{13}$ cm. 
The values of constants 
are given in the Table~\ref{table1}. 
For comparison, we have taken the initial mass and angular momentum of the disk to be the same in both solutions. 
We see that the numerical solution settles onto the analytical solution after the initial transient stage. 
The mismatch near the inner boundary, in the left panel, is a characteristic issue with the self-similar solutions which usually can not accommodate boundary conditions. Nevertheless, this is not important for calculating the mass flow rate at the inner radius.   
The panel on the right shows the mass flow rate at the inner disk which agrees well with the numerical solution
after the brief transient stage.

\placefigure{fig_1}

The early papers studying fallback disks for AXPs \citep{CHN,EkA03,PHN}
employed the dynamical timescale ($\sim$ 1 ms on the surface of the neutron star) 
for $t_0$ in equation~(\ref{Mdot}). 
Numerical calculations of \citet{pringle91} show that power-law evolution starts in a viscous time-scale. 
Comparing analytical and numerical solutions, 
we find  that the 
characteristic viscous timescale $t_0$ of the disk is 
determined by the initial mass 
{\em and} the total angular momentum of the disk. The equations (\ref{scalings}) can be combined to give
\begin{equation}
t_{0}=\frac{4( 4\pi \gamma_2)^{q}
( \gamma_2/\gamma_1)^{4q-2p+4}}{3C(GM)^{2q-p+2}}\frac{J_0^{4q-2p+4}}{%
M_0^{5q-2p+4}}.
\end{equation}
For an electron scattering dominated disk,
\begin{equation}
t_{0}\simeq 10\,\allowbreak \mathrm{y}\,\allowbreak \left( \frac{j_{0}}{%
10^{19}\mathrm{cm}^{2}\mathrm{s}^{-1}}\right) ^{7/3}\left( \frac{M_{0}}{%
10^{-4}M_{\odot }}\right) ^{-2/3}
\end{equation}
showing that the viscous timescale in the disk is very sensitive to the average value of specific angular momentum $j_0$ in the disk. Indeed, $t_0$ increases to $\sim$ 2000 y if $j_0\sim 10^{20}$ cm$^2$ s$^{-1}$, a realistic range according to our calculations in \S\ 3.

For a given initial mass of the disk, setting $t_0$ 
corresponds to setting the initial angular momentum of the disk as the fourth parameter in addition to the initial mass of the disk together with the magnetic dipole moment and the initial period of the neutron star.

The analytical solution presented in this section does not address transient behavior due to thermal-viscous instabilities and/or the variation of the active disk radius. In \S\ 3,  we will explain how these instabilities are 
introduced into the framework.

\begin{figure}[t]
\epsscale{1.0} 
\plottwo{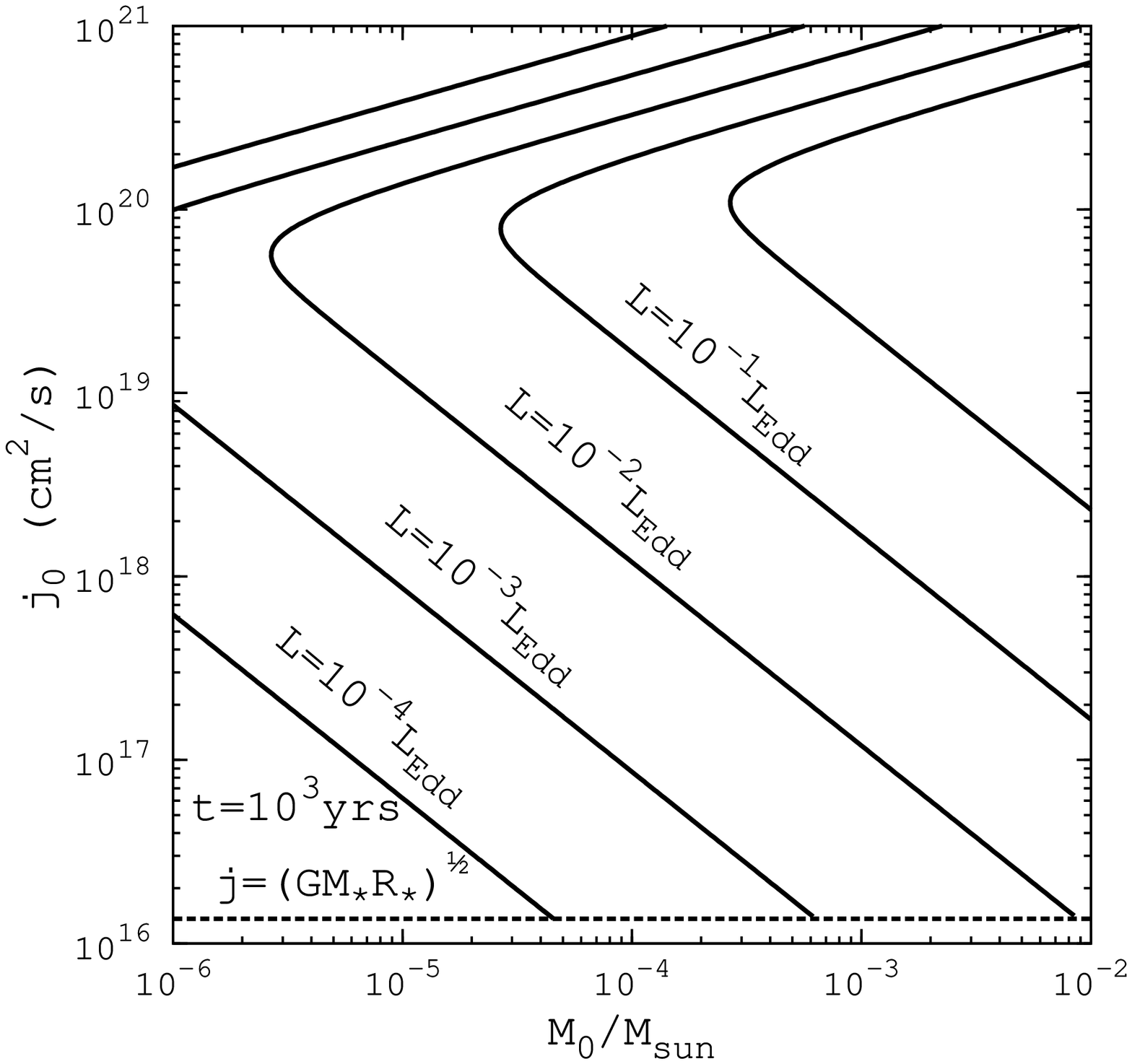}{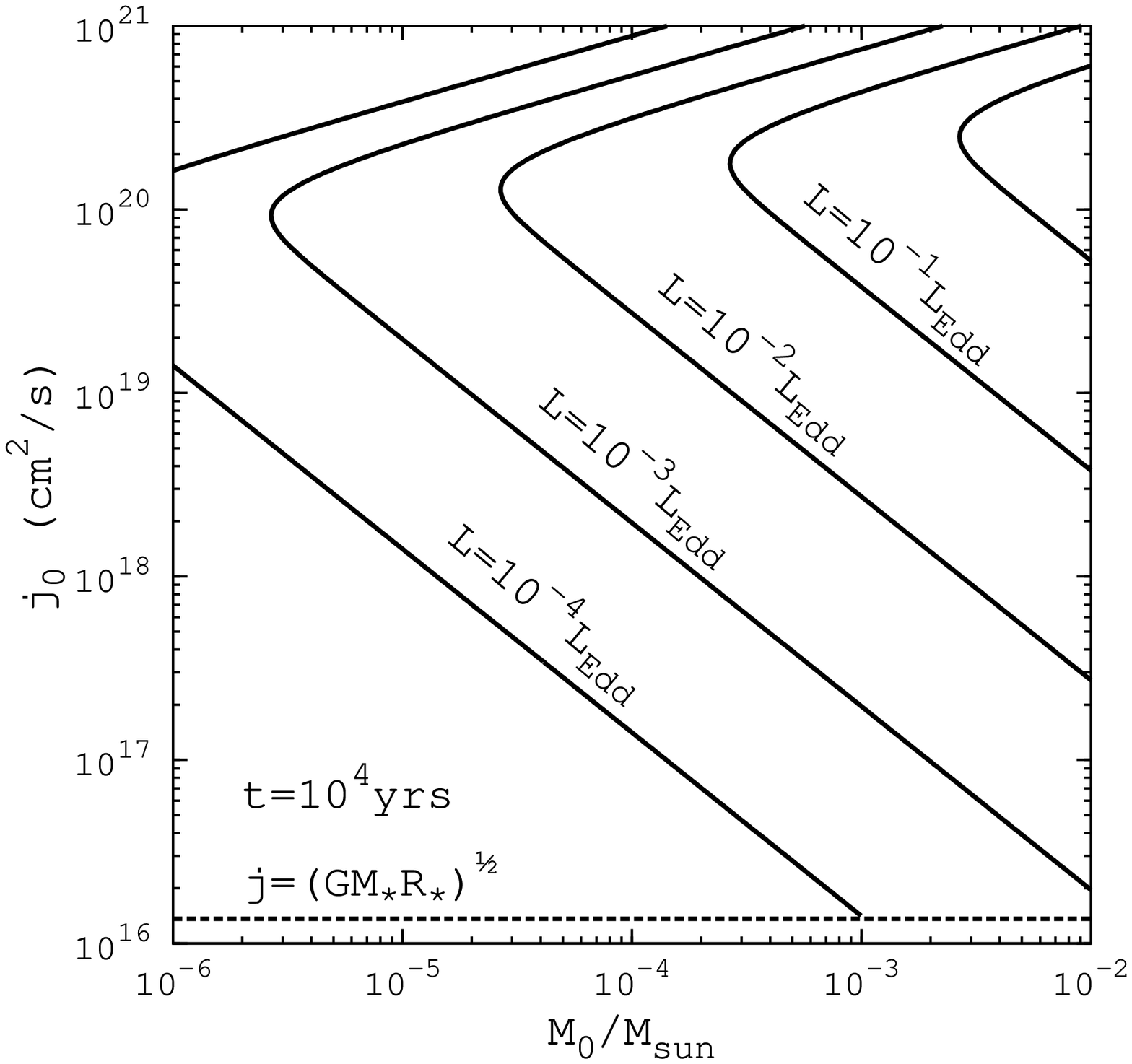}
\caption{The X-ray luminosity of accretion from a fallback disk for a range of initial disk masses and initial specific angular momentum at $t= 10^3$ y (left panel) and at $t=10^4$ y (right panel).}
\label{isoluminos}
\end{figure}

We use the analytical model as a guide to the time consuming numerical calculations, for the choice of initial conditions and to exhibit the resulting neutron star luminosities as shown in Figure~\ref{isoluminos}. 
This Figure shows the iso-luminosity lines of fallback disks at the age of $10^3$ y and $10^4$ y, respectively, depending on the initial mass and initial specific angular momentum of the disk. It is seen that there is a wide band of this parameter space that can lead to a disk evolution with $L=10^{-3}-10^{-2}~ L_{\mathrm E}$, the observed AXP luminosity range, at the age of $10^3-10^4$ y.

\placefigure{isoluminos}

In the following section, for our numerical simulations, we will 
choose an initial disk density distribution corresponding to 
a specific angular momentum of $\sim 9.7 \times 10^{19}$ cm$^2$ s$^{-1}$ 
which leads to AXP luminosities 
with an initial disk mass of $10^{-4}-10^{-5}\Msun$ as can be seen from Figure~\ref{isoluminos}.
These values lead to viscous timescales of $t_0 \sim 2000$ y and $t_0 \sim 9000$ y, 
respectively. These values for $t_0$ are orders of magnitude greater than the dynamical timescale employed in earlier models \citep{EkA03,MPH01}, 
and lead to a slower decline of the disk mass and $\Mdot$. The inner radius of the disk, determined by $\Mdot$, will also be evolving on this long timescale in which the period of the neutron star evolves. This leads to an evolutionary picture  very different from the earlier studies. In earlier work, with ad hoc choices of short $t_0$, the inner radius of the disk was seen to move outwards very rapidly as the disk evolved. The inner disk radius could be kept inside the light cylinder only if the neutron star was subject to large torques, i.e. if it had large magnetic fields ($0.5 - 1\times 10^{13}$ G) so that the star spun-down sufficiently fast and the light cylinder expanded rapidly enough. The torque equilibrium for such high magnetic fields requires high mass flow rates. This led to high luminosities in model calculations. To avoid this problem, a rather efficient propeller was required allowing only $\sim 10^{-2}$ of the mass inflow to accrete onto the neutron star.   
In the present model the realistic slow evolution of the disk allows neutron stars with smaller magnetic fields to become AXP/SGRs.

\section{MODEL CALCULATIONS}

We solve the disk diffusion equation (\ref{diffusion_eq}) for the surface density $\Sigma$ of the disk (see e.g. \citet{FKR02}) to examine the evolution of the X-ray luminosity and the spin period of the neutron star together.  We take the initial surface density distribution as $\Sigma \p r^{-\lambda}$ with  $\lambda = 3/4$ which is the expected surface-density profile of a thin, steady-state accretion disk. Independently from  the actual initial mass distribution of a fallback disk, the disk's subsequent evolution quickly leads to an extended thin disk. 

Opacities depend on the details of the disk composition that are not well known for fallback disks. As in our earlier work on enhancement light curves of AXPs and SGRs, we use electron-scattering opacity in our calculations. Using bound-free opacities does not significantly affect our results. In any case the uncertainty in opacity is folded in with the uncertainty inherent in the $\a$ parameter of the viscosity. We solve the disk diffusion equation by employing the $\a$ prescription of the kinematic viscosity $\nu = \a \cs h$ \citep{SS73} where $\cs$ is the local sound speed and $h$ is the pressure scale height of the disk. 

What should be the value of the $\a$ parameter to be used in our numerical calculations?  To answer this question, we pursue the results of our earlier work on the enhancement phases of persistent and transient AXP/SGRs. X-ray and infrared enhancement light curves of  both persistent and transient AXP/SGRs can be explained by the evolution of the disk after the inner disk matter has been swept  outwards by a soft gamma-ray flare \citep{ErA03,EGA06,EE08}. There is an  important difference between the X-ray outburst light curves of persistent and transient AXPs and SGRs. 
For the persistent AXP/SGRs, model fits are obtained by using a single viscosity parameter ($\a\simeq 0.1$) that is similar to those employed in modeling the hot states of soft X-ray transients (SXTs) and dwarf novae (DN) in their outburst phases. This probably indicates that for persistent AXP/SGRs, the inner part of the fallback disk within a sufficiently large radius remains in the same viscosity state during the enhancement. Fitting the X-ray and IR enhancement light curves from persistent AXP/SGRs  does not give information about whether the disk is  in a colder (lower) viscosity state at the outer regions of a fallback disk. Help comes from the transient AXP/SGRs whose low luminosity quiescent phases and transient bright phases indicate transitions between different viscosity states of the active disk. The light curves of the transient sources can be reproduced if the disk  undergoes a thermal-viscous disk instability at a critical temperature $\Tcrit \sim 1000 - 2000$ K  \citep{EE08,ErCa08}. As temperatures drop below $\Tcrit$ the viscosity parameter changes from $\ah \simeq 0.1$ to
$\ac \simeq 0.025 -0.035$ in exactly the same way applied in the disk instability models of SXTs and DN for which  $\Tcrit \sim 10^4$ K corresponds to the ionization temperature of hydrogen (see e.g. \citet{Lasota01} for a review of the disk instability model).  
This difference between transient and persistent AXPs and SGRs could be attributed to their different accretion rates prior to the enhancement. In the long term, it is the cold state viscosity that drives the mass-flow rate from the outer to the inner disk in quiescence. The viscosity parameter $\ah$ of the hot inner disk does not significantly change the long-term evolution of the disk. In all model calculations, we take $\ac = 0.03$, $\ah = 0.1$, and $\Tcrit = 1000$ K. We note that in both hot and cold states the disk is active, providing mass inflow with turbulent viscosities. The disk instabilities at $\Tcrit \sim 1000$ K do not affect the long-term evolution. Using only a single $\a =\ac$, without introducing a $\Tcrit$ in the model, the same results obtained for the long term evolution of the disk are almost the same as in the case of a two-state disk. 

Our main concern here is on how the long-term evolutions of AXPs and SGRs are affected if the disk becomes passive below some temperature $\Tp$ which is much lower than and should not be confused with $\Tcrit$. Termination of the disk's active lifetime by transition to a final passive phase at very low temperatures $\Tp$ seems to characterize the $\Mdot$ and $P$ evolution of the AXPs and SGRs in the long term.

The outer disk of a neutron star in X-ray binaries are cut by tidal forces. 
There are no such forces that can constrain the extension of a fallback disk around AXP/SGRs. 
Where is the outer radius $\rout$ of a fallback disk?  
By this $\rout$ we mean the radius beyond which active viscous mass transport ceases. 
A strong possibility is that the outermost disk could be passive at very low temperatures. 
 To put the question in other words, 
what is the minimum temperature below which disk regions enter a passive (non-turbulent) 
low viscosity regime? The work by \citet{IS05} shows that the viscosities are expected to 
be turbulent by the magneto-rotational instability (MRI) (Balbus \& Hawley 1991), even at temperatures $\sim$ 300 K. In agreement with this result, 
in the present work, we show that a rather low minimum active disk temperature $\Tp \sim 100$ K, 
below which the viscosities are not turbulent, can explain the period and  X-ray luminosity 
evolution of AXPs and SGRs consistently with the observed period, period-derivative and 
X-ray luminosity distributions of these sources. 

The minimum active disk temperature, $\Tp$, should not be confused with the critical temperature, 
$\Tcrit$, leading to a disk instability that was proposed by \citet{EE08} as the mechanism responsible for the 
X-ray enhancement of the transient AXP XTE J1810-197. Unlike $\Tp$, $\Tcrit$ has no significant effect 
on the long-term evolution of a fallback disk. It determines the transition between cold and 
hot state viscosities that governs the mass-flow rate, $\dot{M}$, in the active parts of the disk.

The position of $\rout$ is not important in simulating the short-term events like enhancements lasting from a few months to a few years as long as $\rout$ is taken sufficiently large in order to avoid numerical effects due to artificial cutoff of the disk. Numerically the outer disk must be cut somewhere. A constant $\rout$ is not realistic for a fallback disk and causes numerical effects on the model $\Mdot$ evolution in the long term. The natural transition to a passive disk, occurring at ~$T = \Tp$, determines $\rout$ as the radius beyond which ~$T < \Tp$. Thus, $\rout$ is a dynamical quantity, moving from the outer disk inward as the disk ages and cools.

From the model fits to the X-ray enhancement data of the transient AXPs and SGRs, we  find that fallback disks are likely to be active even at temperatures about 1000 K. At much lower temperatures it is plausible to expect that a fallback disk gradually becomes passive starting from the outer radius which has the lowest temperature along the disk.  At radii larger than some transition radius $\rp$, at temperatures lower than a critical temperature $\Tp$, the outermost disk becomes passive.  As the mass accretion rate decreases with time, leading to decreasing X-ray irradiation strength, the passive outermost disk grows ($\rp$ moves inwards). The transition temperature $\Tp$ below which a fallback disk becomes locally passive is of crucial interest for the late evolution of the disk.

We keep the inner disk radius $\rin$ constant. In the numerical model, separation of the radial grid points decreases with decreasing radial distance $r$ (see \citet{ErA03}).  Narrower radial grid separation requires smaller viscous time step. Therefore, time steps for the calculations along the radial grids become  smaller with decreasing inner disk radius. Due to these numerical reasons, we take $\rin =10^{10}$ cm which is larger than the expected inner disk radius of a fallback disk around an AXP with a conventional dipole magnetic field of strength $B_{\ast }\sim 10^{12}$ G on the surface of the neutron star. This simplification does not affect the mass-flow rate at the inner disk, $\Mdotin$. The realistic inner disk radius of the disk is expected to be close to the Alfv\'{e}n radius, 
$\rA =\Mdotin^{-2/7}(G M_{\ast })^{-1/7}\mu _{\ast }^{4/7}$, where $\mu_{\ast }$ is the dipole magnetic moment and  $M_{\ast }$ is the mass of the neutron star. 
We calculate the disk torque acting on the neutron star through the interaction of the inner disk with the magnetic dipole field of the neutron star 
using $\mu _{\ast }= 10^{30}$ G  together with the values of $\Mdotin$ and $\rA$ calculated at each time step (see \citet{EE08} for details). 
Thus, the fastness parameter $\fast$ and the torque are calculated dynamically, using the current $\rA$ value at each step. The fixed choice of $\rin$, for numerical reasons, therefore does not influence the torque evolution either.

The torque model is based on the angular momentum exchange between the
stellar magnetosphere and the disk matter that penetrates through the
effective magnetic coupling radius $r_{\mathrm{A}}$ in the fast rotator
regime with $\omega _{\ast }>1$, where $\omega _{\ast }=\left( r_{\mathrm{A}%
}/r_{\mathrm{co}}\right) ^{3/2}$ is the fastness parameter. We assume that
the effective magnetic coupling radius is given by the Alfv\'{e}n radius $r_{%
\mathrm{A}}$. Here, the co-rotation radius, $r_{\mathrm{co}}=\left( GM_{\ast
}/\Omega _{\ast }^{2}\right) ^{1/3}$, represents the critical radius where
the disk matter rotates with the same speed as the neutron star. This is
true if the rotation of the disk is purely Keplerian. In reality, the actual
rotation rate of the disk matter slightly exceeds the local Keplerian value
at each radius $r$ between the effective co-rotation radius and the Alfv\'{e}%
n radius. This is the result of the transfer of angular momentum from the
neutron star to the inner portion of the disk threaded by the stellar
magnetic field lines over the range $r_{\mathrm{co}}\lesssim r\leq r_{%
\mathrm{A}}$. In this regime of spin-down with accretion, the magnetic stresses
gradually force the inner disk matter to co-rotate with the neutron star
within a transition region in the disk where the rotation rate $\Omega $ 
changes from
nearly Keplerian values in the outer parts to co-rotation with the magnetosphere at the inner boundary of the disk.

For a magnetically coupled disk-neutron star system, the integration of
magnetic stresses over the magnetically threaded region of the disk yields
the effective magnetospheric torque on the neutron star (see \citet{ErA04}). 
In the mildly fast rotator regime ($\omega _{\ast}\gtrsim 1$), assuming most of the
incoming disk matter, instead of being propelled out of the system, accretes
onto the neutron star, the effective magnetospheric torque can be estimated
as%
\begin{equation}
N_{\ast }\simeq -\int_{r_{\mathrm{co}}}^{r_{\mathrm{A}}}r^{2}B_{\phi
}^{+}B_{z}dr=\frac{\gamma _{\phi }}{3}\dot{M}\sqrt{GMr_{\mathrm{A}}}\left(
1-\omega _{\ast }^{2}\right) ,  \label{mgtrq}
\end{equation}%
where $B_{\phi }^{+}=\gamma _{\phi }B_{z}$ is the azimuthal component of the
magnetic field on the surface of the disk, $\gamma _{\phi }>0$ is the
azimuthal pitch of order unity for $r\gtrsim r_{\mathrm{co}}$, and $%
B_{z}\simeq -\mu _{\ast }/r^{3}$ is the vertical component of the stellar
dipole field threading the inner disk over the range $r_{\mathrm{co}%
}\lesssim r\leq r_{\mathrm{A}}$ \citep{EE08}.

The disk torque in equation (\ref{mgtrq}) has been recently used by \citet{EE08} to explain the spin-frequency evolution of the transient AXP
XTE J$1810-197$ during the long-term post-burst fading of its X-ray flux. The
disk model has also accounted for the X-ray light curve of the AXP XTE J$%
1810-197$ during the same period (see \citet{EE08}). The model fits
to both the X-ray light curve and spin-frequency data are consistent with
the regime in which the neutron star spins down while accreting most of the
matter flowing in through the disk. When the inner-disk radius is outside the
 light cylinder, we assume in the torque calculations that the inner disk extends down to the light 
cylinder  and that all the mass is propelled out 
of the system. In the model calculations, this happens during the early 
evolutionary phases, and these assumptions do not change the subsequent 
evolution curves of the model sources.     

As in the case of the AXP XTE J$1810-197$, X-ray luminosity changes in 
the transient AXPs and SGRs cover
a broad range from $\sim 10^{35}$ erg s$^{-1}$ down to
$\sim 10^{33}$ erg s$^{-1}$. Such a wide range of X-ray luminosities, in a
given transient source, encompasses both the relatively high accretion
regime ($\sim 10^{14}-10^{16}$ g s$^{-1}$) of the persistent AXPs and SGRs
and the low accretion regime ($\lesssim 10^{13}$ g s$^{-1}$) in the
quiescent phase of their transient cousins. As shown by \citet{EE08}, 
the disk torque in equation (\ref{mgtrq}) describes well the spin
evolution of the neutron star during the long-term X-ray enhancement
of the transient AXP XTE J$1810-197$ while the X-ray flux changes by
up to two orders of magnitude. The dependence of the torque on the fastness
parameter, that is, $N_{\ast }\propto 1-\omega _{\ast }^{2}$, accounts for
the spin-frequency evolution of the relatively fast rotating ($\omega _{\ast
}>1$) neutron stars for which the long-term decay of the X-ray luminosities
can be explained by a decrease in the mass-inflow rate in the inner disk
(see, e.g., \citet{EE08}). The same torque, on the other hand,
behaves as $N_{\ast }\propto 1-\omega _{\ast }$ for the neutron stars that
are close to rotational equilibrium with the disk, i.e., for $\omega _{\ast
}\simeq 1$ (see, e.g., \citet{EEEA07}). Whether they are close to
rotational equilibrium or not, the torque in equation (\ref{mgtrq}) is
applicable to both the persistent and transient sources and thus will be
employed in our calculations for the evolution of the spin-period and X-ray
luminosities of the AXPs and SGRs.

\section{RESULTS AND DISCUSSION }

In our calculations, we see that for a given initial disk mass, $M_0$,   
and a minimum active disk temperature $\Tp$, there is a minimum initial period 
$\P0min$ such that sources with initial period $P_0 < \P0min$  cannot evolve 
into the AXP phase. For the sources with $P_0 > \P0min$, time for the onset 
of accretion depends on $P_0$. The neutron stars having  longer $P_0$ start 
accreting matter earlier. For a large range of $M_0$, all the model 
sources that can enter the regime of accretion with spin-down converge into
the AXP/SGR phase. When the initial mass of the disk is too large, above a certain $M_0$, X-ray luminosities remain 
higher than  $10^{36}$ erg s$^{-1}$ (or equivalently $\Mdot > 10^{16}$ g s$^{-1}$) 
for sufficiently long time intervals, which would make sources with $\Lx > 10^{36}$ erg s$^{-1}$. Since, such luminuous sources are not observed, there is probably an upper limit to $M_0$ values.

\begin{figure}
%\figurenum{2a}
\epsscale{1.42}
\hspace{-2.5 cm}
\plottwo{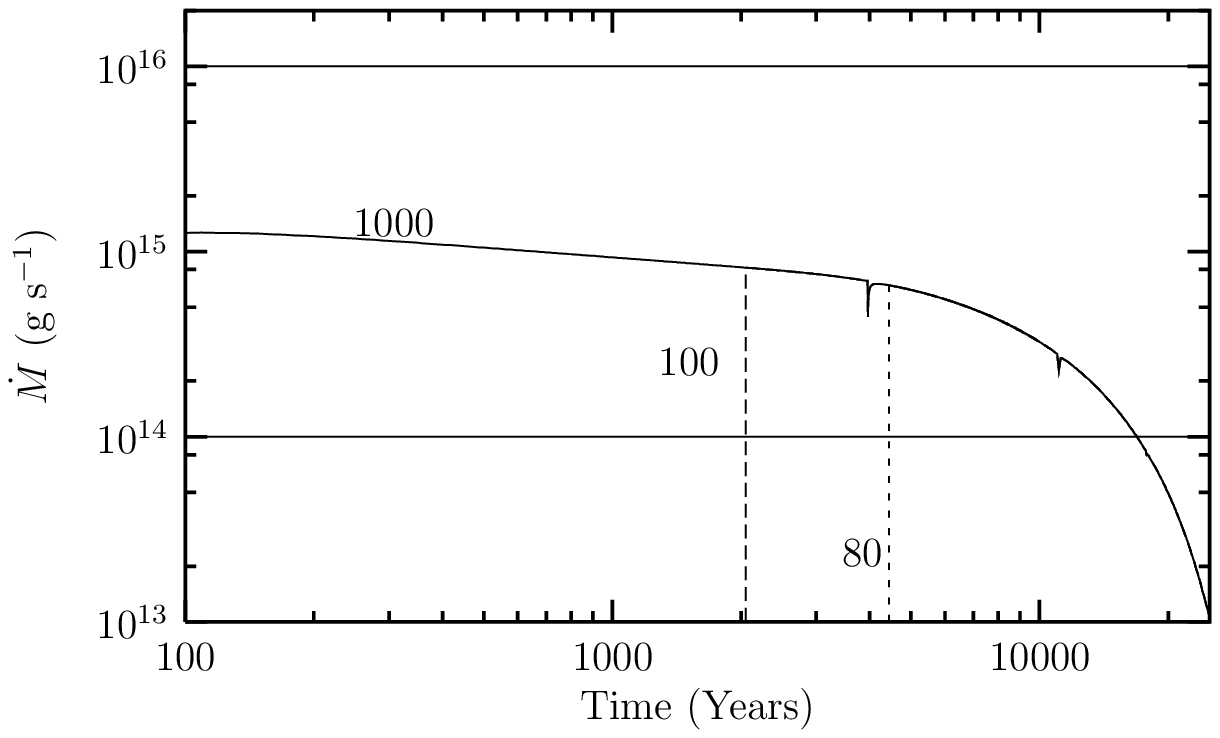}{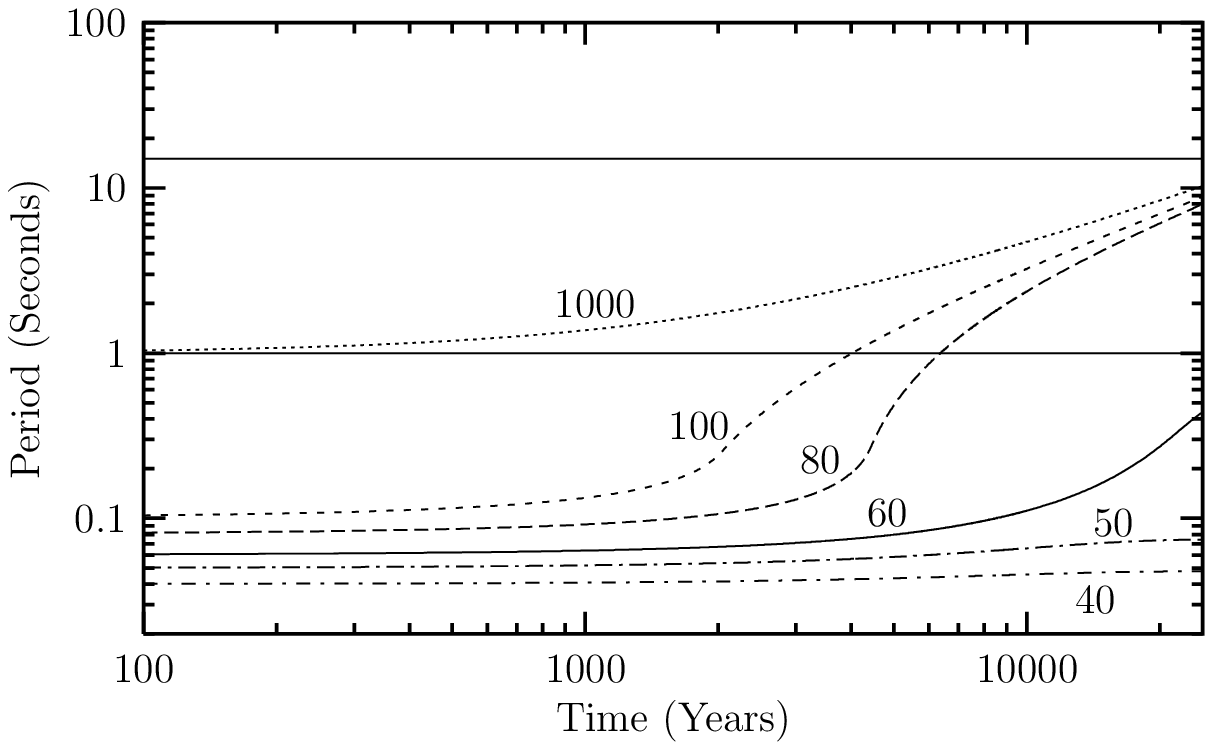}
%\vspace{-12.5 cm}
\caption{(Left panel) $\Mdot$ evolution with initial mass $M_0 = 1 \times 10^ {-5} \Msun$ of the extended disk and with the transition temperature between active and passive viscosity states $\Tp= 100$ K.  The models with the shown initial periods (in ms) start on the accretion phase of the same evolutionary track at the times indicated with dashed vertical lines. For this $M_0$, sources with $P_0$ less than about $\sim 70$ ms are likely to evolve as radio pulsars (see the text for details). The two small spikes result from local viscous instabilities in the disk and are not numeric artifacts. (Right panel) Period evolution with the same parameters given in the left panel. Model curves with $P_0$ values of 80, 100 and 1000 ms (noted on the left panel) are seen to enter the AXP/SGR phase.  The sources with  initial periods 40, 50 and 60 ms do not reach the spin-down with accretion regime.}
\label{fig2}
\end{figure}

\begin{figure}
\epsscale{1.42}
%\figurenum{3a}
%\vspace{-11 cm}
\hspace{-2.5 cm}
\plottwo{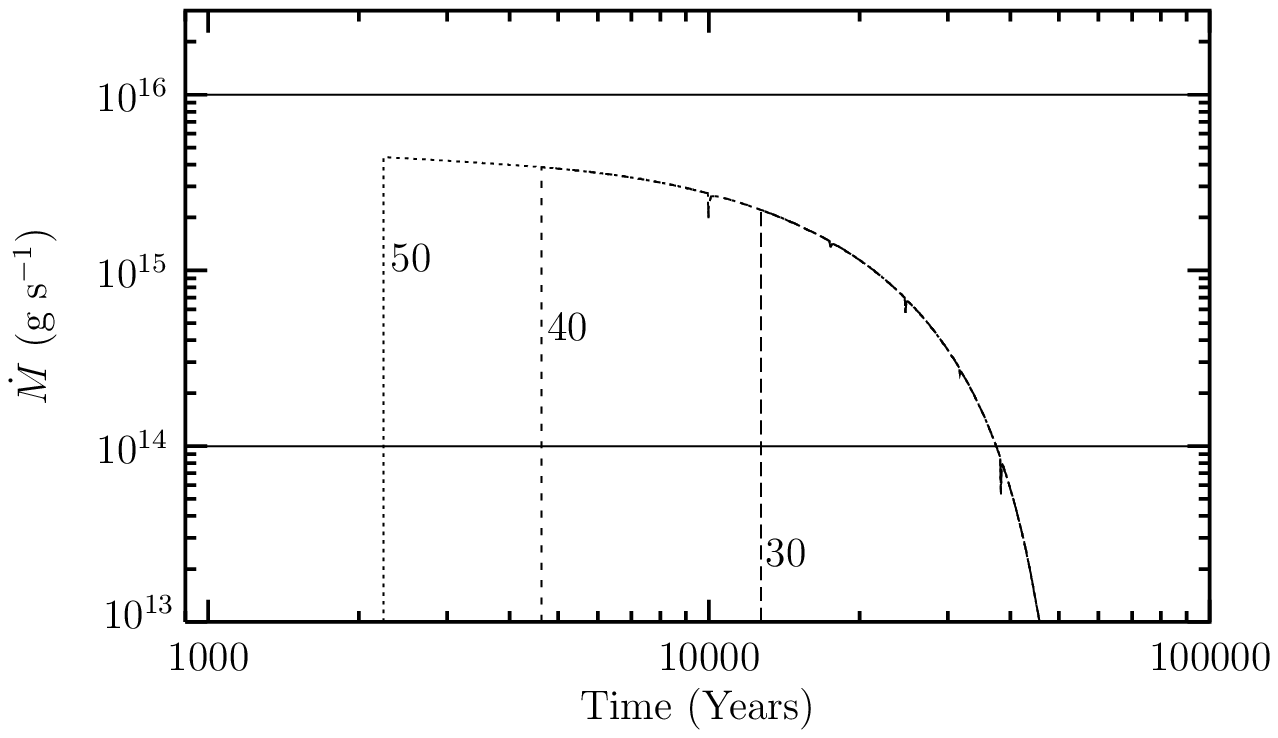}{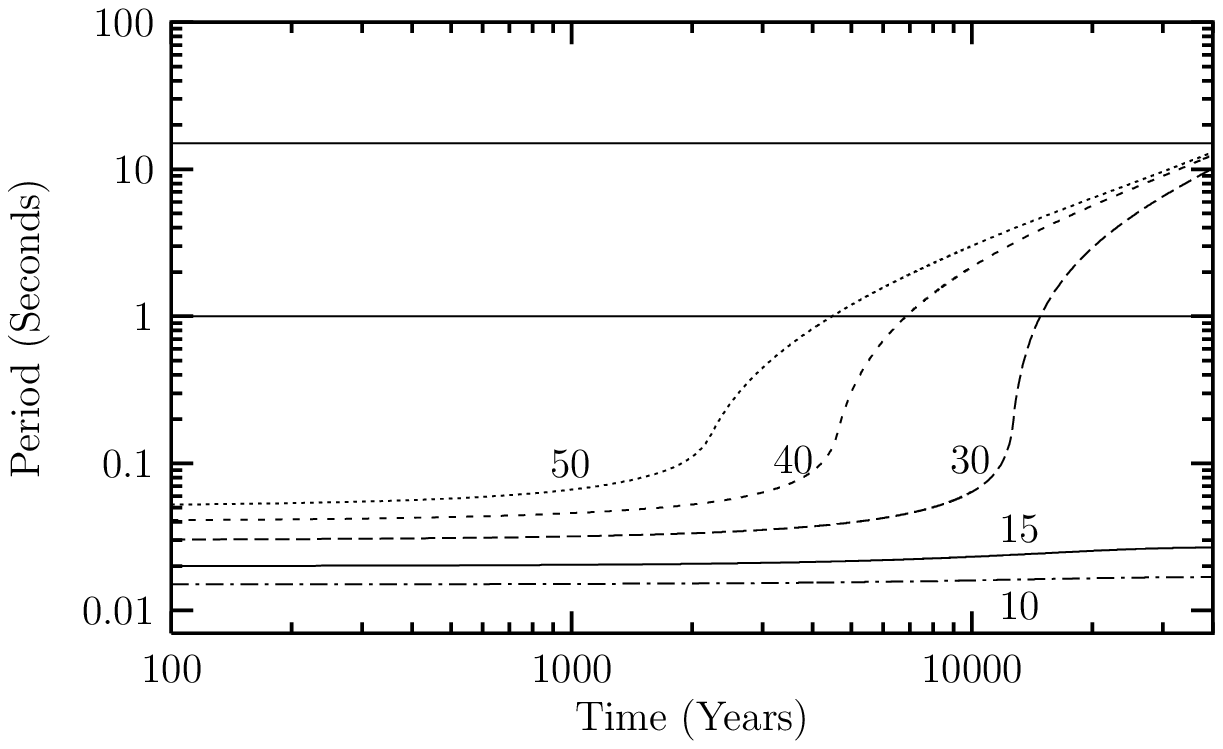}
%\vspace{-13 cm}
\caption{(Left panel) Same model as given in Fig.~\ref{fig2} with $M_0 = 4 \times 10^ {-5} \Msun$, and initial periods  30, 40 and 50 ms. (Right panel) Period evolution for the same model with $\Mdot$ curves seen in the left panel. With this initial disk mass, the model sources having initial periods 10 and 15 ms cannot become AXP/SGR.}
\label{fig3}
\end{figure}

\begin{figure}
\epsscale{1.42}
%\figurenum{4a}
%\vspace{-11 cm}
\hspace{-2.5 cm}
\plottwo{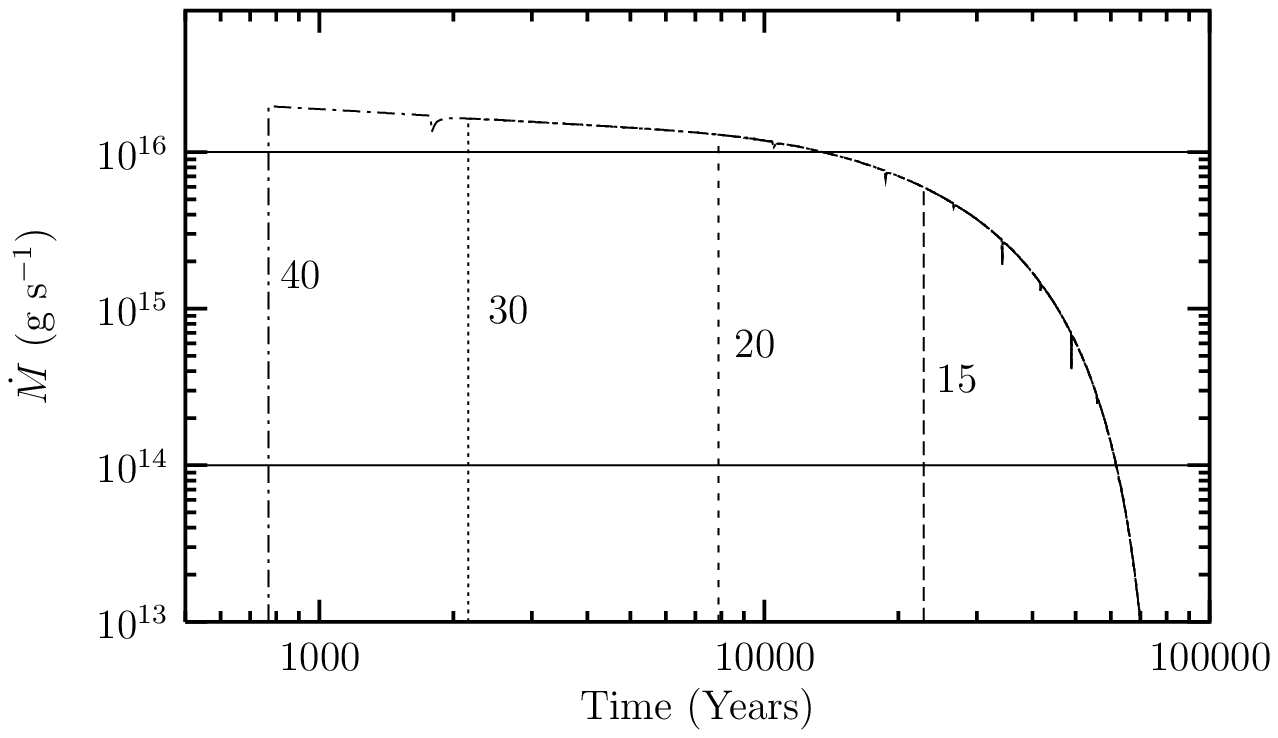}{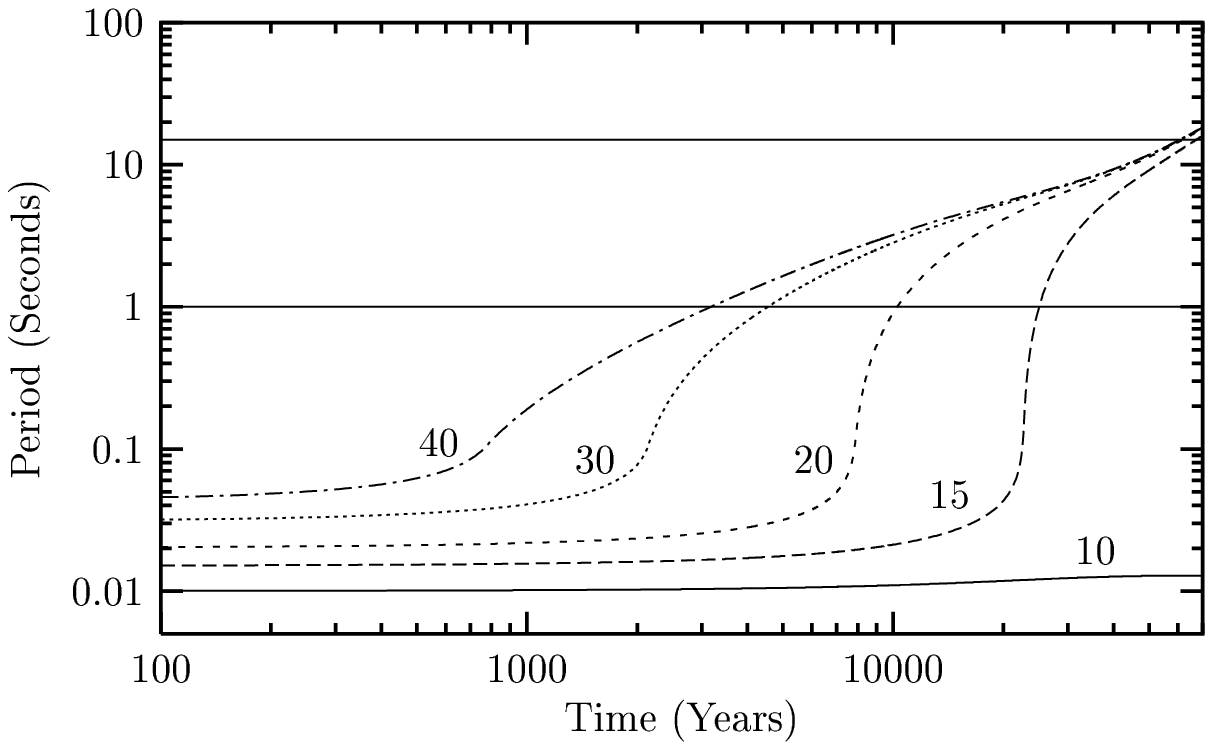}
%\vspace{-13 cm}
\caption{(Left panel) The same model as given in Fig.~\ref{fig2}, with $M_0 = 1 \times 10^{-4} \Msun$. Initial periods are  15, 20, 30 and 40 ms.; (Right panel) Period evolution corresponding to the $\Mdot$ curves given in the left panel.  
The source having initial period of 10 ms cannot become an AXP for this initial mass,  
while all the other sources, comparing with left panel, 
are seen to evolve into AXP phase.}
\label{fig4}
\end{figure}

\begin{figure}
\epsscale{1.42}
%\figurenum{5a}
%\vspace{-11 cm}
\hspace{-2.5 cm}
\plottwo{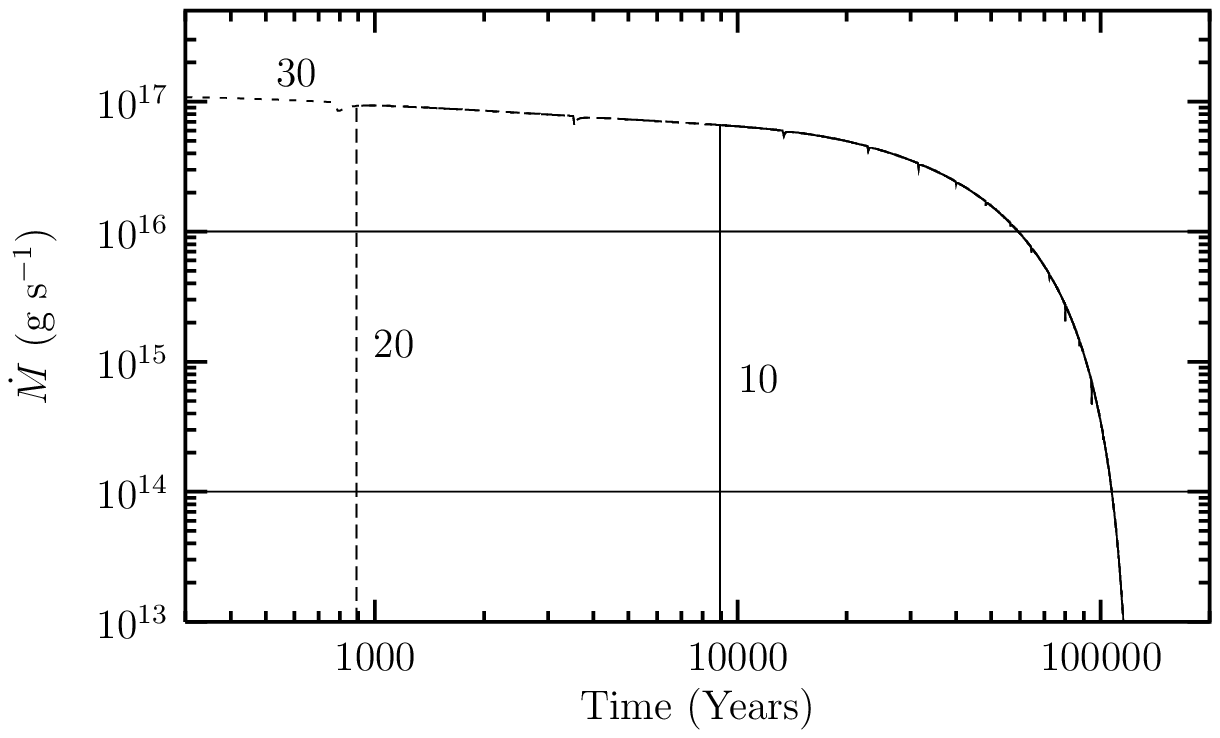}{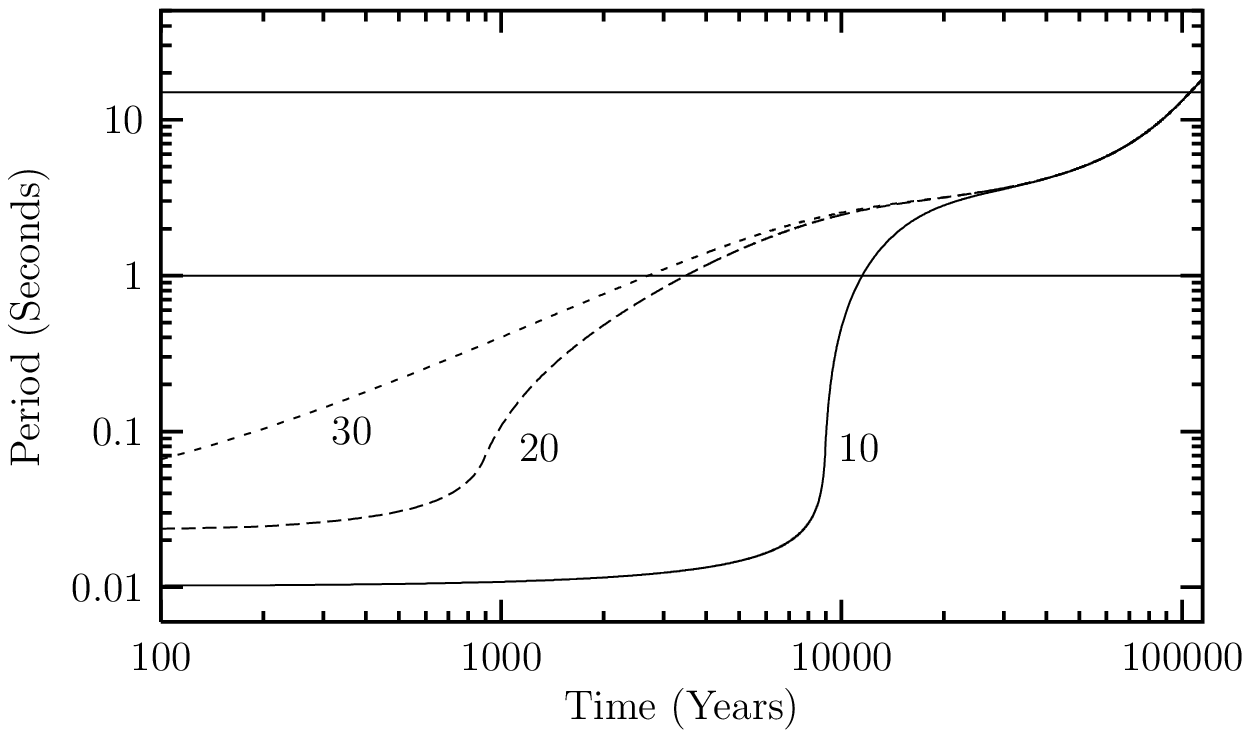}
%\vspace{-12.5 cm}
\caption{(Left panel) The same model as given in Fig.~\ref{fig2} except $M_0 = 3.5 \times 10^{-4} \Msun$. Initial periods are  10, 20 and  30 ms. This initial mass  seems to be unlikely for fallback disks  unless a large fraction of the inflowing disk matter is propelled out of the system for an initial evolutionary phase of a few $\times 10^4$ y. Otherwise, we would expect to see sources with luminosities above the observed AXP luminosities and with periods around or less than the AXP periods (see the right panel). (Right panel) Period evolution corresponding to the $\Mdot$ curves given in the left panel.}
\label{fig5}
\end{figure}

In the first set of simulations (Figs. \ref{fig2}$-$\ref{fig5}), we set $\Tp = 100$ K. For different values of initial total disk mass $M_0$, we obtain $\Mdot$
and period evolution curves. In the figures we have indicated the lines 
for $\dot{M}=10^{14}$ g s$^{-1}$ and $\dot{M}=10^{16}$ g s$^{-1}$ to delineate the $\dot{M}$ interval
corresponding to the observed AXP/SGRs.       
In Figure~\ref{fig2}, it is seen that for $M_0 = 1 \times 10^{-5} \Msun$  
 only the sources  with $P_0 > 70$ s can evolve to the AXP phase, 
and those with lower $P_0$ are likely to become, and remain as radio pulsars  throughout their 
lifetimes.  Model sources with initial periods 80, 100, and 1000 ms  start 
accretion at different times, while all of them evolve into the AXP phase. The minimum initial period required for the source to eventually enter the AXP phase,  $\P0min$, is correlated with the initial mass of the fallback disk;   
$\P0min$ decreases with increasing $M_0$.  For $M_0$ values of 
$4\times 10^{-5}, 1.0\times 10^{-4}$ and $3.5\times 10^{-4} \Msun$ 
we obtain minimum initial periods around 30 , 15 and less than 10 ms respectively 
(Figs. \ref{fig3} - \ref{fig5}). 
For a large range of $M_0$ values, Figures \ref{fig2} - \ref{fig5} clearly show that the sources with  $P_0 > \P0min$  converge to the AXP/SGR phase and more significantly, their period evolution converges to, and remains in the period range of AXP/SGRs as their luminosities decrease below minimum AXP luminosities.
Given the observed range of X-ray luminosities, the model curves seen in Figure 6 are likely to represent the evolution with an extremely high $M_0$ that is probably either unrealistic or not common. It is remarkable that, even in this case, $\Mdot$ and $P$ curves tend to meet in the AXP/SGR phase.

\begin{figure}
\epsscale{1.42}
%\figurenum{6a}
%\vspace{-11 cm}
\hspace{-2.5 cm}
\plottwo{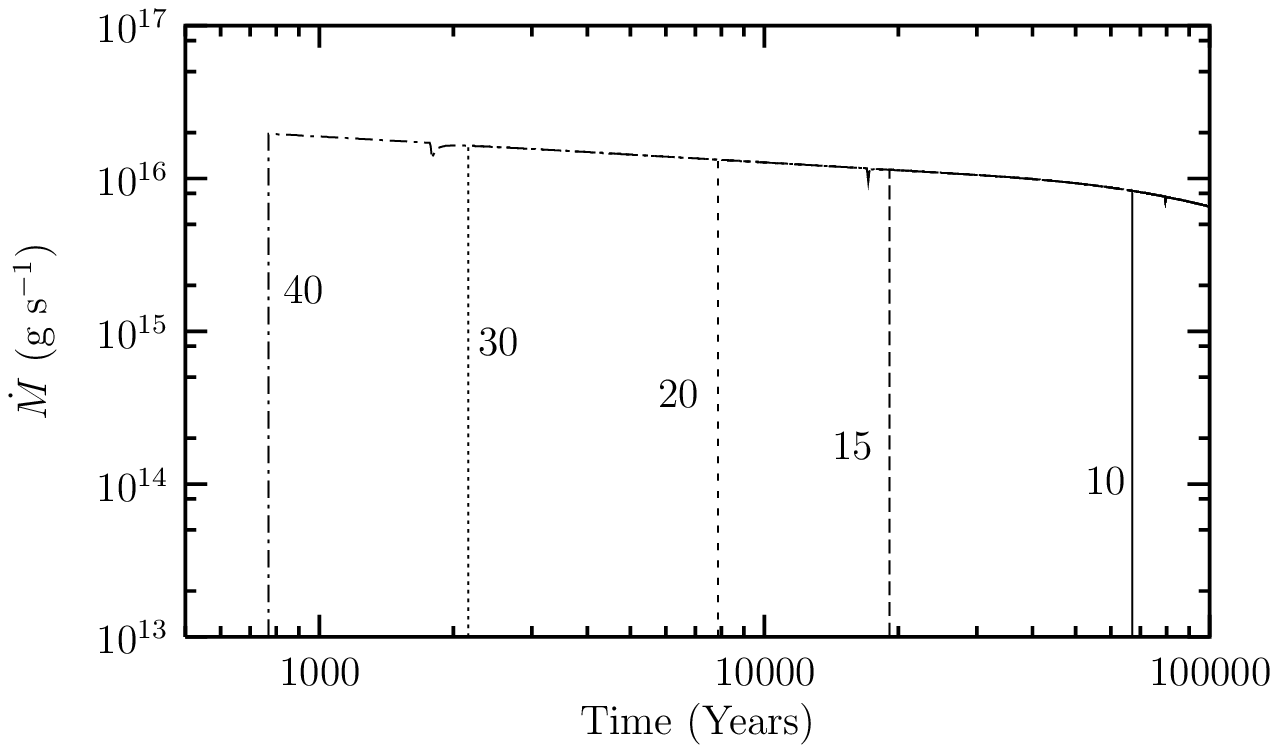}{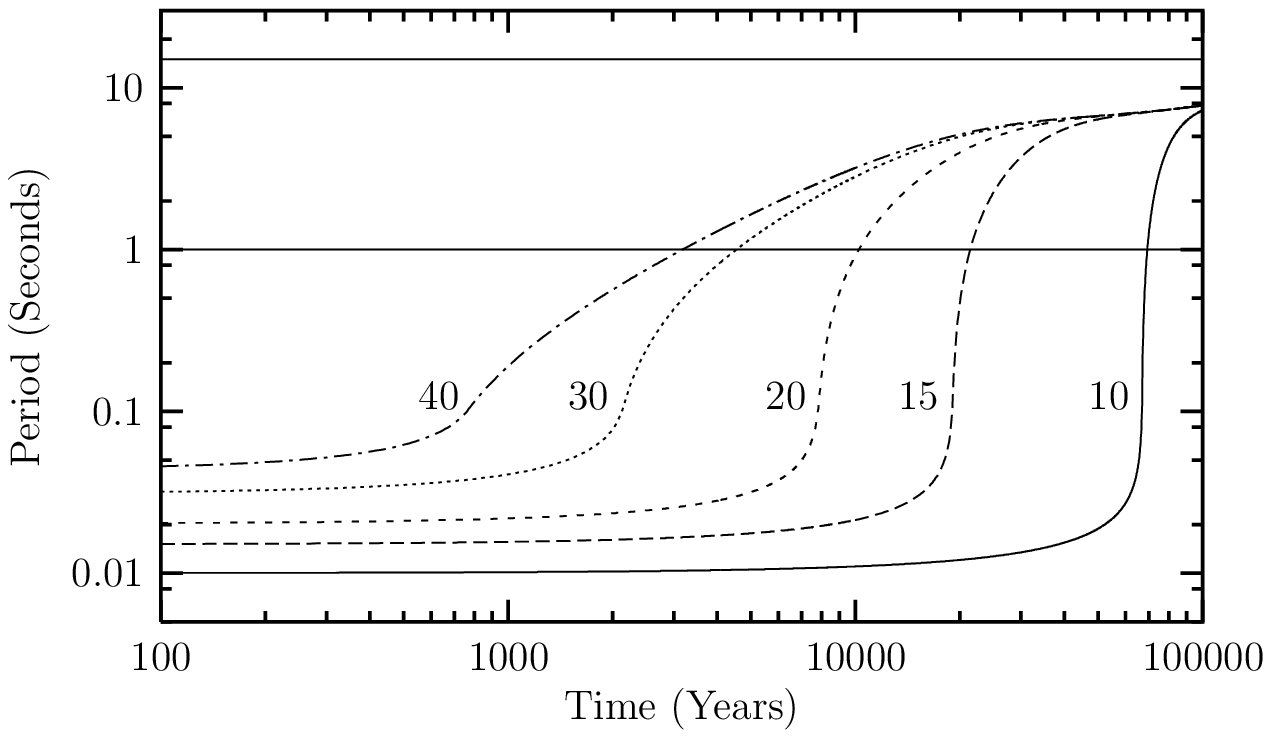}
%\vspace{-13 cm}
\caption{(Left panel) The only difference from the model given in Fig.~\ref{fig4} is $\Tp = 0$, that is, the entire disk is taken to be active throughout the calculations. The model $\Mdot$ curves shown in the figure are obtained for different initial spin periods: 10, 15, 20, 30 and  40 ms. (Right panel) Period evolution for the same model  model given in 
the left panel.} 
\label{fig6}
\end{figure}

How do these results change if we set $\Tp =0$, that is, if the disk remains active for all temperatures? This is actually an implicit assumption in all earlier analytical work using self-similar evolution of  fallback disks \citep{MPH01,EkA03}. For this test,  we set $M_0= 1\times 10^{-4} \Msun$ (same as for the models given 
in Fig.~\ref{fig4}) and $\Tp = 0$ without changing the other parameters. We repeat the calculations for different initial periods.  The resultant model curves are presented in Figure~\ref{fig6}. It is seen in these figures  that model sources starting with different neutron star initial periods do settle into the AXP/SGR period and luminosity range. As expected, there is a crucial difference between the models with $\Tp = 0$ K and $\Tp = 100$ K in that the cutoffs in $\Mdot$ curves for $\Tp = 100$ K are not seen in $\Mdot$ evolutions with $\Tp = 0$. This means that the mass-inflow rate remains large, above or slightly below $10^{16}$ g s$^{-1}$, for $10^5$ years or more. Thus, without a cutoff in disk activity, agreement with the observed AXP/SGR luminosity range would require an explanation. Ek\c{s}i \& Alpar (2003) suggested that a large fraction of mass flowing in through the disk is not actually accreted onto the neutron star, requiring rather efficient propeller activity.    

\begin{figure}
\epsscale{1.42}
%\figurenum{7a}
%\vspace{-13 cm}
\hspace{-2.5 cm}
\plottwo{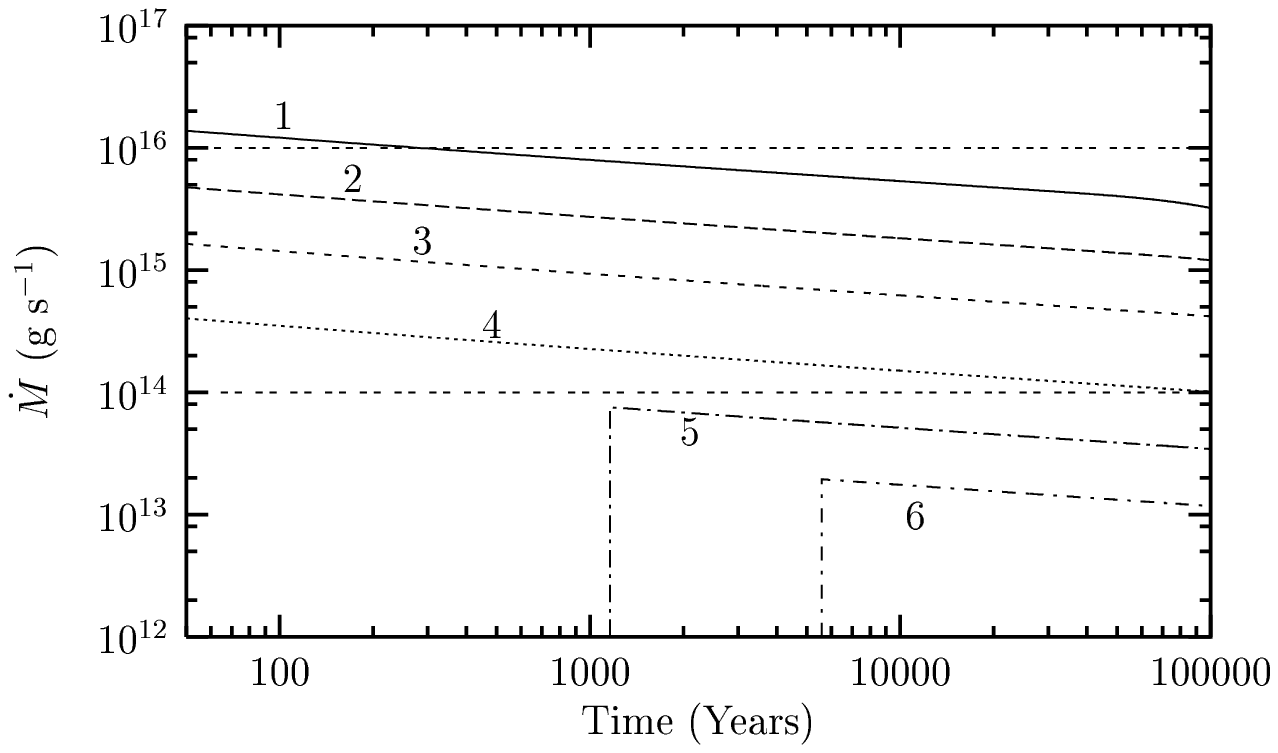}{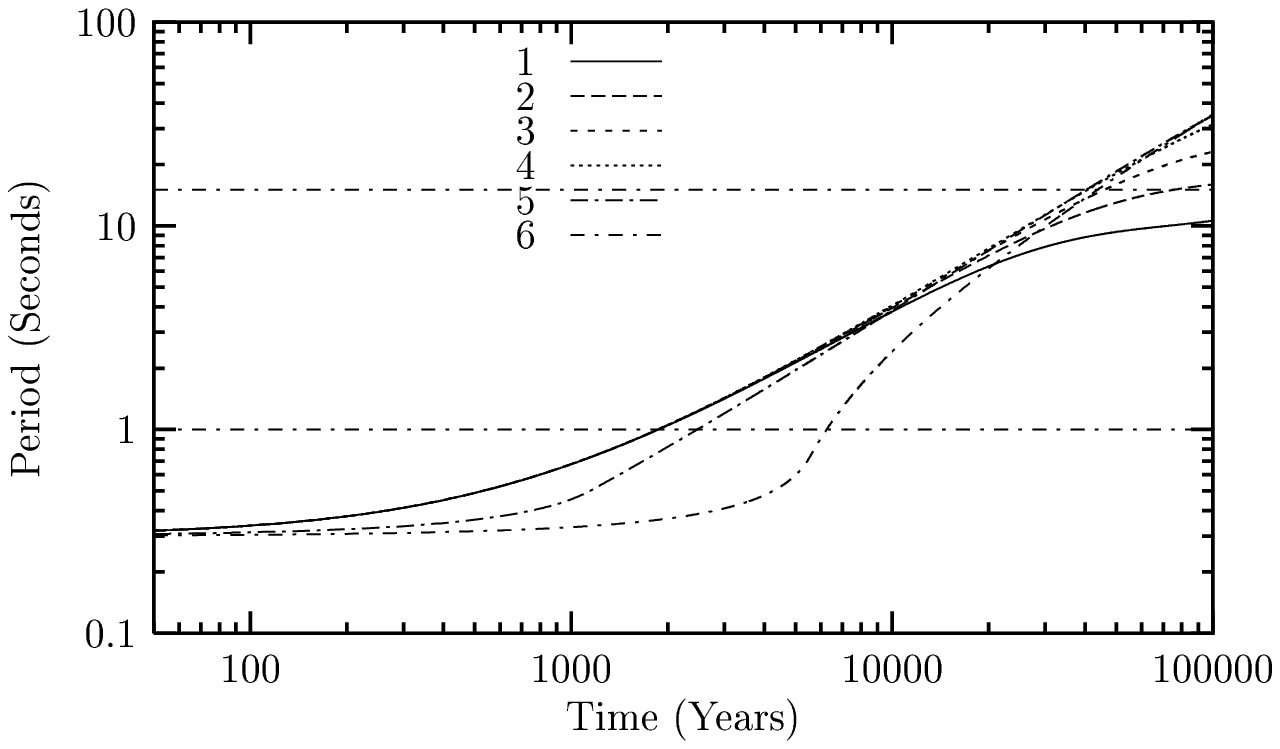}
%\vspace{-12.5 cm}
\caption{(Left panel) $\Mdot$ evolution for $\Tp = 0$ with different $M_0$ values: $1\times 10^{-4 }\Msun$ (1), $5.8\times 10^{-5 }\Msun$ (2),  $2.5\times 10^{-5} \Msun$ (3), $3.5\times 10^{-6} \Msun$ (4), $1.6\times 10^{-6} \Msun$ (5), and $7.5\times 10^{-7}  \Msun$ (6). For a comparison of the model curves we take 
$P_0 = 300$ ms for all the models (see the text). (Right panel) Period evolutions accompanying the $\Mdot$ evolutions given in the left panel. Period clustering is also expected for $\Tp = 0$. There is an important difference between the $\Mdot$ evolution curves of the models with $\Tp =0$ and $\Tp= 100$ K (see Figure~\ref{fig8}). For $\Tp = 0$, we would expect to see many AXPs and SGRs  with ages greater than $10^5$ y and with periods even larger than observed AXP and SGR periods (see the text for discussion).} 
\label{fig7}
\end{figure}

We have calculated more models  for a better comparison of evolution with and without the disk becoming eventually passive. First, we take $P_0 = 300$ ms, $\Tp = 0$ and repeat the calculations for different $M_0$ values (Fig. \ref{fig7}). (We note that the  model sources with low $M_0$ values (curves 5 and 6 in Figure~\ref{fig7}) that are seen to evolve towards the AXP phase with $P_0 = 300$ ms,  would become radio pulsars for initial periods $P_0 \lesssim 100$ s). This shows that sources with a large range of initial periods evolve into the AXP range once they have started to accrete matter. 
The model curves in Figure \ref{fig7} correspond to different initial disk masses ranging from $7.5\times 10^{-7}$ to $1\times 10^{-4} \Msun$. All these illustrative models remain within the range of AXP/SGR luminosities for more than $10^5$ y. During most of this epoch, the sources lie in the AXP/SGR period range as well. Nevertheless, if this model were a correct representation of these systems we would expect to see many more, mostly older, AXPs and SGRs with similar luminosities and with periods larger than the observed  range of period clustering. This is a serious difficulty for fallback disk models without an eventual cutoff for $\dot{M}$. 

Next, we do the same exercise with the same $P_0$ and the same set of $M_0$ values, but for $\Tp = 100$ K 
(see Figure~\ref{fig8}). The model sources that enter the period range of AXP/SGRs follow almost the same period-evolution curve. They enter the AXP phase and remain there  for a long time interval lasting from a few $\times 10^3$ y to a few $\times 10^4$ y  depending on the initial mass of the fallback disk. Subsequently, periods of the sources tend to increase with about the same rate to beyond the maximum observed AXP periods. By comparing panels a and b in Figure~\ref{fig8}, we see that the sudden turn down  in X-ray luminosity renders the sources unobservable  at an age of about a few $10^4$ y. This terminates the observable period evolution in an epoch when the periods are clustered in the $2-12$ s range, in agreement with observations.

For given initial disk mass $M_0$,  how does the evolution depend on $\Tp$? 
To illustrate, we compare the $\Mdot$ and period evolution with $\Tp$ values of 0, 60, 80, 100 and 200 K for $M_0= 1 \times 10^4 \Msun$ and $P_0 = 30$ ms. 
For this $M_0$, the model sources having $\Tp$ greater than about 250 K cannot evolve into the AXP phase. This is because higher $\Tp$ means earlier  decrease in mass-flow rate from the outer to the inner disk, and thus the inner disk cannot penetrate into the light cylinder (see Figure~\ref{fig9}). Taking also the earlier model fits with different disk masses into account, the results obtained with $\Tp \sim 100$ K are in good agreement with observations of AXPs and SGRs.

In our model calculations, we take  $B=10^{12}$ G.  
For larger dipole magnetic fields of the neutron star the eventual, approximately equilibrium periods reached in the late evolutionary epochs would be longer in proportion to $B^{6/7}$. To maintain agreement with the observed period  clustering requires earlier luminosity cutoff. 
In the model, for larger dipole magnetic fields of the neutron star  up to about $1 \times 10^{13}$ G,  it is still possible to reproduce AXPs by increasing $\Tp$ and $M_0$ values accordingly. Increasing  $B$ results in a narrower initial $M_0$  range that can give AXPs. This is because increasing field strength requires higher $M_0$ for the inner disk to be able to penetrate the light-cylinder. For the field strengths above $\sim 10^{13}$ G, model sources cannot enter the AXP phase, both the luminosity and the period derivative remain well above the observed values while the neutron star is tracing the AXP period range. 
Note that the $B$-field discussed here is the neutron star surface value of the 
{\em dipole} component, as this is the long range component of the magnetic field that determines the torques and interactions with the disk.   
The magnetar fields of AXPs and SGRs are likely to be in higher multipoles in view of the crustal field amplification mechanisms responsible for magnetar strength fields. It is the dipole fields that must have magnitudes  less than about $10^{13}$ G. Recent work on AXP surface spectra \citep{GOG07,GOG08} has produced estimates of the total surface magnetic field strengths, in AXPs XTE~1810$-$197 and 4U~0142+61. The total surface fields inferred were $\sim 3 \times 10^{14}$ G and $\sim 5 \times 10^{14}$ G respectively, larger than the $B \lesssim 10^{13}$ G requirement of our model for the dipole component alone. 
Our result also implies that some of the young neutron stars that have  high dipole fields  $B > 10^{13}$ G are not likely to evolve into the AXP phase even if there are fallback disks around them.

This paper does not address the relative strengths of dipole and higher multipole components on the basis of neutron star physics. We do think, however, that dipole fields weaker than the higher multipole fields are plausible for magnetar models. The large fields in the star's crust are amplified, rearranged in crust breaking events and dissipated in the neutron star crust. Magnetic stresses are here in interaction with crustal crystal stresses, which have critical strains, dislocation and domain wall distributions that introduce local scales to the problem. In the neutron star crust with very efficient screening, the crystal is close to a Coulomb lattice. So the relevant scales of microcrystalline alignment can be much longer than those in terrestrial crystals, but still they are small compared to the global size of the neutron star. Field generation and reconnection in coupling with discrete crust breaking effects will plausibly put energy into higher multipoles \citep{ruderman91}. 

In $P, \Pdot$  and timing signatures, there is a continuity between AXPs, SGRs and the radio pulsars. If the torque causing the spin-down is the dipole spin-down torque for an isolated neutron star, than the dipole magnetic field is a function of $P$ and $\Pdot$ ( assuming all neutron stars in the sample have standard values of $R_{*}^6/ I$; possible effects of variations of this were discussed by \citet{guseinov05}).  Then the neutron stars with similar positions in the $P-\Pdot$ diagram would be continuous in $B_{*,\mathrm{dipole}}$ also, leaving a puzzle as to why some are AXPs and SGRs and others not.  Furthermore, if the dipole spindown torque is operating on all AXPs, SGRs and high $B_{*,\mathrm{dipole}}$ pulsars, evolutionary connection between AXP/SGRs and high $B_{*,\mathrm{dipole}}$  radio pulsars would require ongoing dipole field generation, as constant field evolutionary tracks for isolated neutron stars have slope -1 in the $\log \Pdot - \log P$ diagram. 

Presence or absence of fallback disks and their properties bring in a new parameter, not determined by $P$ and $\Pdot$ alone, into play. The dipole fields inferred by using fallback disk model spindown torques are somewhat lower than the dipole fields inferred using the isolated rotating dipole model. The transformation from $\Pdot$ and $P$ to $B_{*,\mathrm{dipole}}$ involves the mass inflow rate $\Mdot$ and fastness parameter, whose values may be constrained by observations of the X-ray luminosity for some sources. Discussion of the $P-\Pdot$ diagram in the light of this transformation will be the subject of our subsequent work.  In any case, sources located close to each other in the  $P-\Pdot$ diagram are not likely to have similar values of $B_{*,\mathrm{dipole}}$  if their spindown is due to fallback disk torques. Inferred dipole fields of a radio pulsar and an AXP lying close to each other in the $P-\Pdot$ diagram will be different also in the case that the radio pulsar does not have a fallback disk, so it has a relatively higher $B_{*,\mathrm{dipole}}$  inferred by the dipole spindown torque, compared to the AXP which will have a relatively lower $B_{*,\mathrm{dipole}}$  inferred by the disk spindown torque.  

If there are radio pulsars with fallback disks, their surface dipole magnetic fields, calculated correctly with the disk torques are likely to be in the $B_{*,\mathrm{dipole}} \sim 10^{12}$ G range. These can evolve to become AXPs. Our results show that pulsars with $B_{*,\mathrm{dipole}} > 10^{13}$ G are not likely to become AXPs.

Why do the high $P \Pdot$   radio pulsars (except one case) not show bursts (do not have higher multipoles? After the dipole fields are calculated with the appropriate disk torque for those neutron stars with fallback disks, there may remain still some radio pulsars with surface dipole fields comparable to those of AXPs. There is also the one example (at present) of a bursting radio pulsar PSR J1846-0258 in SNR Kes 57 \citep{gavriil08}. ~So the question arises as to what makes an exceptional bursting radio pulsar. 
If as we claim the burst behavior requires magnetar-strength fields in higher multipoles due to a history of crust breaking, than the usual, non bursting radio pulsars (in general) do not have strong higher multipoles even if some of them may have rather strong surface dipole fields.  A likely explanation is that magnetar field evolution starts with the initial total magnetic moment of the neutron star, and is driven by the initial field in the neutron star core. Suppose a magnetar field in higher multipoles would obtain if the initial total field in the neutron star's core was higher than some critical value, enough to affect crust breaking (say by the magnetic flux lines extending through the superconducting core). The high dipole field radio pulsars started with fields near or just below the threshold for generating higher multipole magnetar fields: thus the dipole field in these pulsars, like in other radio pulsars is roughly the total field. The one exception, PSR J 1846-0258, may be the one marginally above threshold in its initial neutron star core magnetic moment.

Finally, we compare the $\Pdot$ values of the model sources during their AXP/SGR phase with observations. 
For all models with different $M_0$ and $P_0$ reaching the AXP/SGR  phase, 
we see that $\Pdot$ values remain in the range from a few $\times 10^{-13}$ to $\sim 10^{-11}$ s s$^{-1}$. This is consistent with the observed period derivatives of these sources.  
We present  $\Pdot$ and fastness variations for the same illustrative model seen in Figure~\ref{fig9} with  $\Tp=0$ and 100 K (Fig.~\ref{fig10}a). The corresponding  evolution of the fastness parameter, $\omega _{\ast }$, for the same model is given in Figure~\ref{fig10}b. The maximum of $\omega _{\ast}$ at $t \sim 2000$ y corresponds to the time at which the inner disk penetrates the light cylinder. Before $t \sim 2000$ y, the inner disk remains outside the light cylinder. In this early phase of evolution, during which we take the inner disk radius equal to the light cylinder radius,  the strength of the disk torque increases with increasing light-cylinder radius caused by rapid spin-down of the neutron star. For this example,   after $t \sim 2000$ y, the inner disk remains inside the light-cylinder and the model sources evolve first to the so called tracking phase along which    
$\omega _{\ast }$ stays close to unity, that is, inner disk comes close to the co-rotation radius. 
In Figure~\ref{fig10}b, it is seen that $\omega _{\ast }$ of the source with $\Tp= 100$ K turns up from the tracking phase at about a few $10^{4}$ y  and starts to increase, while the model source with $\Tp= 0$ remains in the tracking phase. This is because the mass-flow rate for $\Tp= 100$ K decreases faster than that for $\Tp=0$ K. The resultant rapid increase in the inner disk radius for   $\Tp= 100$ K causes $\omega _{\ast }$ to increase to higher values. The source is likely to become a strong propeller. A fraction of the mass would be expelled from the system, instead of accreting onto the neutron star. Qualitatively this would further decrease the observed luminosity and make the luminosity turn-off even sharper. Work on the luminosity changes of transient AXPs \citep{EE08} shows that mass expulsion does not yield a self-consistent model for luminosities down to $\sim 10^{33}$ erg s$^{-1}$, in conclusion in agreement with \citet{Rappaport04}. The effect needs to be investigated at lower $\Mdot$  (and higher $\fast$).    

There are several possibilities for the radio emission in AXPs with fallback disks. One of them is the conventional radio-pulsar emission mechanism. It is possible that this mechanism could work in the presence of accretion, with the strong magnetic field confining the accretion column. This could shield and allow the radio emission along the open field lines 
\citep{trumper09}. The accretion column is likely to have a thin cylindrical-shell geometry enclosed by the closed field lines with a radius that depends on the inner disk radius, where the inner disk matter couples to the closed field lines. Accretion energy is injected into the neutron star through the base of the column and radiated from the entire surface of the neutron star in the soft X-ray band, while the column could act as the site of the beamed high-energy emission. In neutron stars with long spin-periods, like the AXPs, the radio emission is expected to be strongly beamed with a full beaming angle of several degrees along the open field lines. This indicates that the radio emission efficiency in these systems is much lower than that estimated by assuming isotropic radio emission. Small beaming angles could explain why the pulsed radio emission is observed from only 2 of the AXPs. 

A second alternative could be the radio emission through gaps provided by the disk-star radio emission mechanism \citep{CR91,EC04}. 
In this picture, the accretion torque, rather than the magnetic dipole torque, powers the gap. 

Another possibility for the radio emission could be related to charge acceleration by magnetar multi-pole fields close to the neutron star as suggested by Thompson (2008) if the site of the radio emission can be shielded from the accretion column.  

Comparing the various radio emission models with the radio and X-ray behavior of the transient AXPs XTE J1810-197 and 1E 1547-54 will be the subject of our future work

\section{CONCLUSIONS}

\begin{figure}
\epsscale{1.42}

%\figurenum{8a}
%\vspace{-15 cm}
\hspace{-2.5 cm}
\plottwo{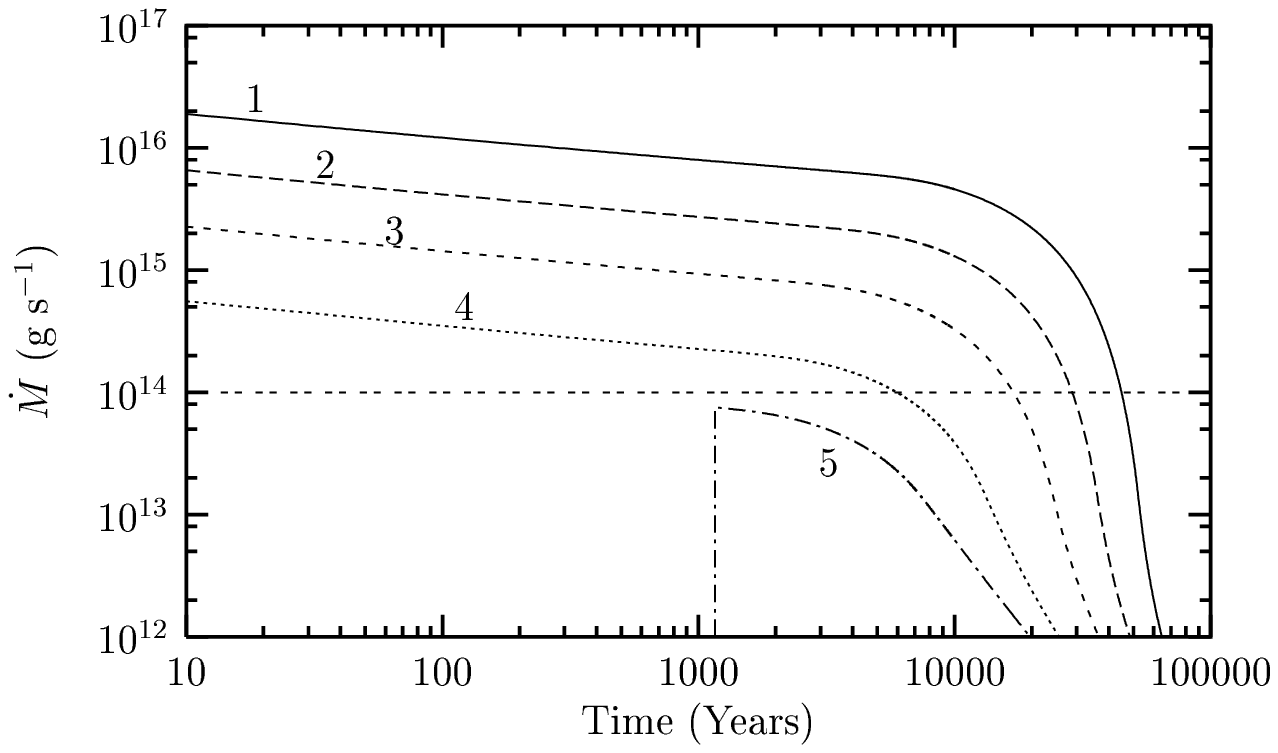}{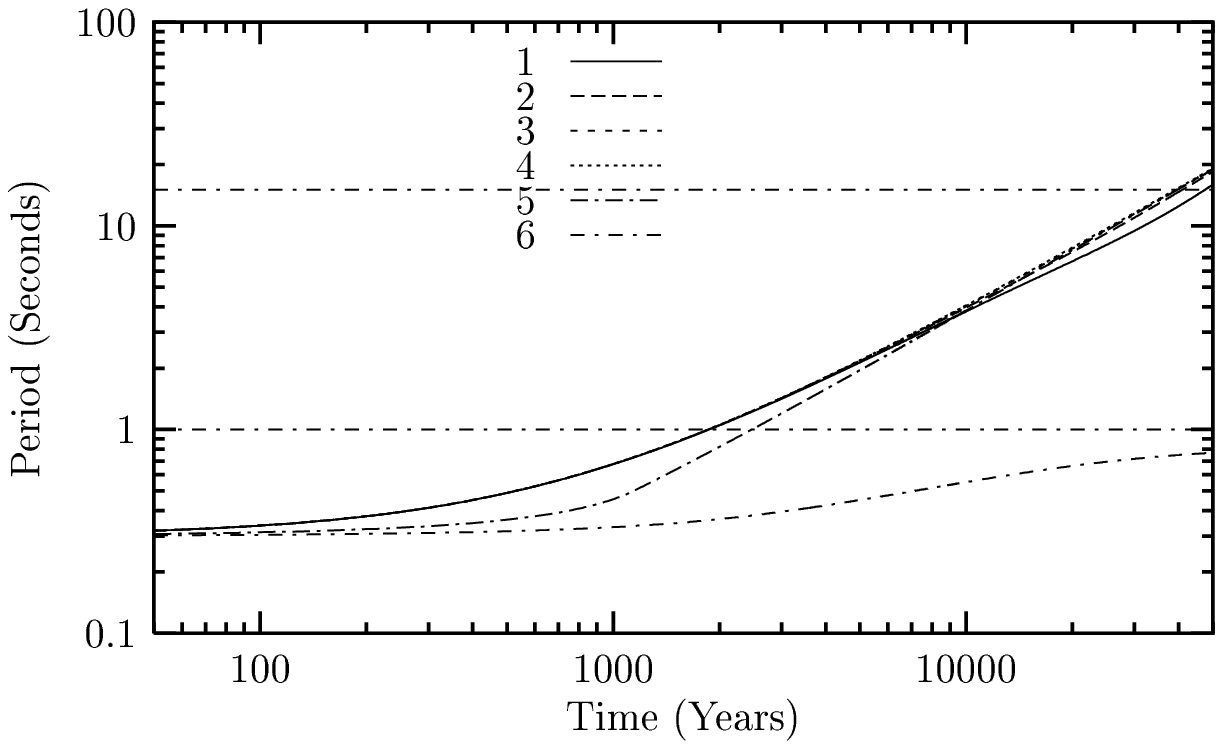}
%\vspace{-12.5 cm}
\caption{(Left panel) The same models as given in Figure~\ref{fig7}, with {\bf $\Tp = 100$ K} here. It is clearly seen that there is a natural cutoff for $\Mdot$ and thus for X-ray luminosity. We see by comparing to right panel that statistical expectation of this model seems to be in good agreement with observations. (Right panel) Period evolution for the model curves given in the left panel. Model (6) with $M_0 = 7.5 \times 10^{-7} \Msun$  is not present in the left panel,  since its inner disk radius cannot penetrate into the light cylinder and the  source evolves as a radio pulsar.  It is striking that all the other sources which enter the spin-down with accretion phase also remain in a long lasting AXP/SGR phase. Furthermore, the model predicts that there will be an upper limit to the observed periods of AXP/SGRs in their present luminosity range because of the luminosity cutoff at the end of the evolution.}
\label{fig8}
\end{figure}

\begin{figure}
\epsscale{1.35}
%\figurenum{9a}
\vspace{4 cm}
\hspace{-2.5 cm}
\plottwo{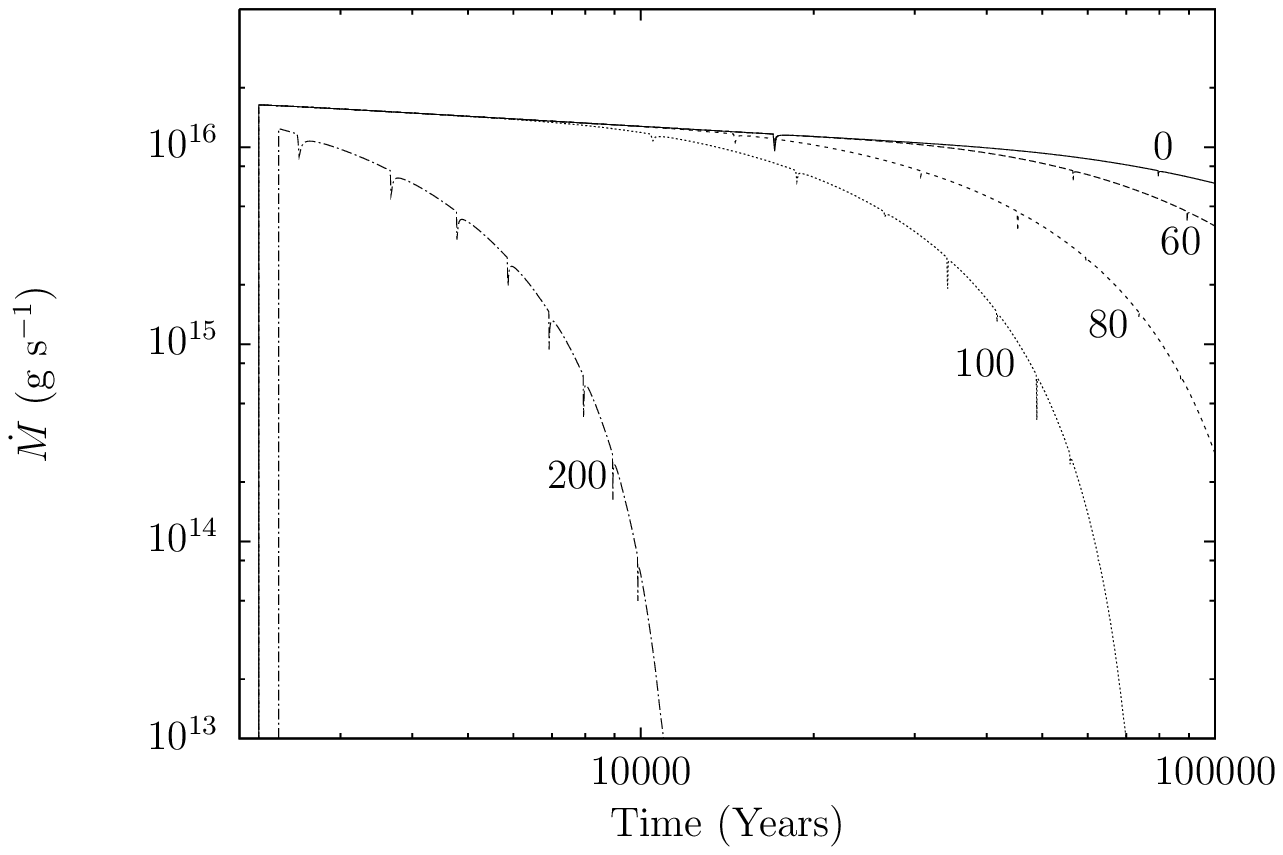}{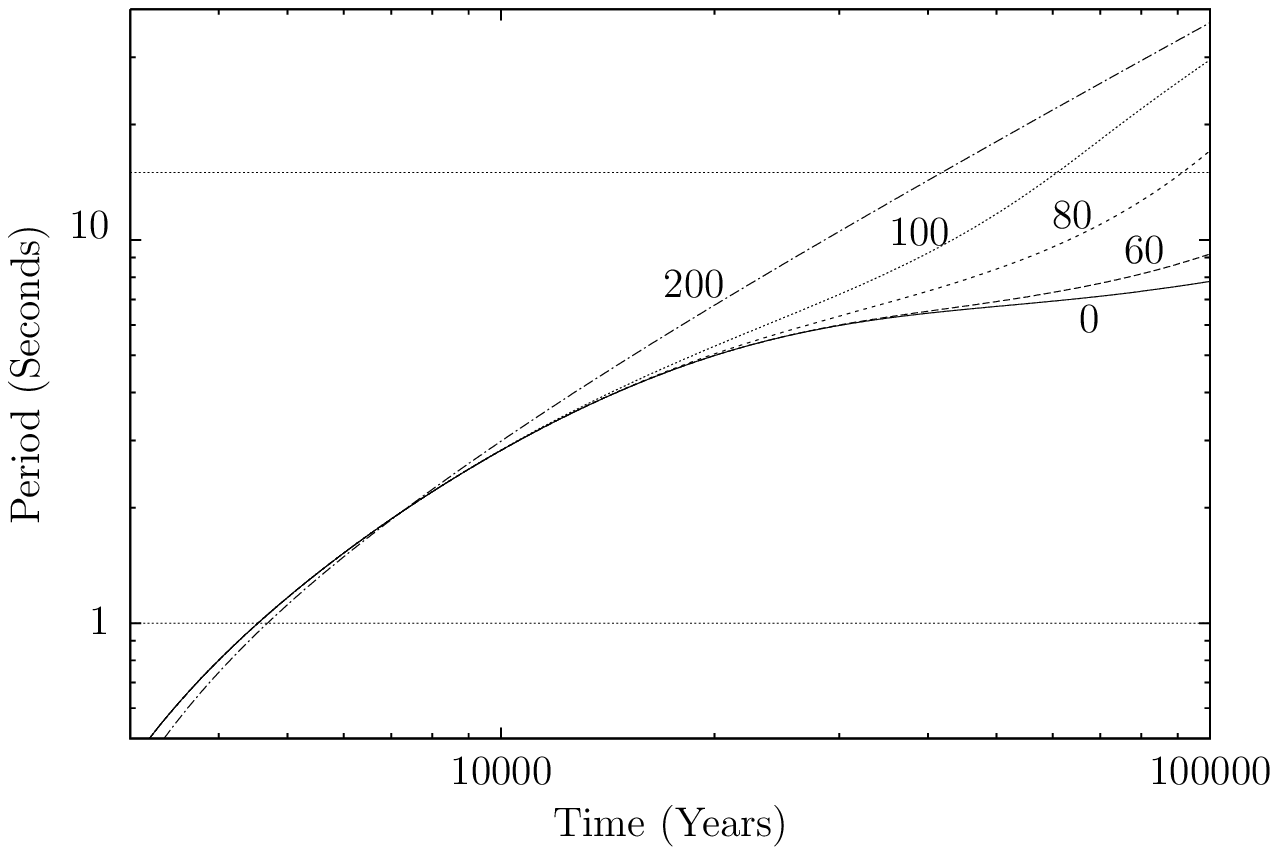}
%\hspace{0.1 cm}

\vspace{-5 cm}
\caption{(Left panel) The same model as presented in Figure~\ref{fig4}, but with different $\Tp$ values which  are seen on the curves in Kelvins. The initial period is 30 ms for all the model sources. With this initial period, the model sources having $\Tp$ greater than about 250 K cannot become an AXP, since higher $\Tp$ values lead to lower mass-flow rate from the outer to the inner disk. (Right panel) Period evolution of the same model sources given on the left panel.}
\label{fig9}
\end{figure}

\begin{figure}
\epsscale{1.45}
%\figurenum{10a}
%\vspace{-13 cm}
\hspace{-2.6 cm}
\plottwo{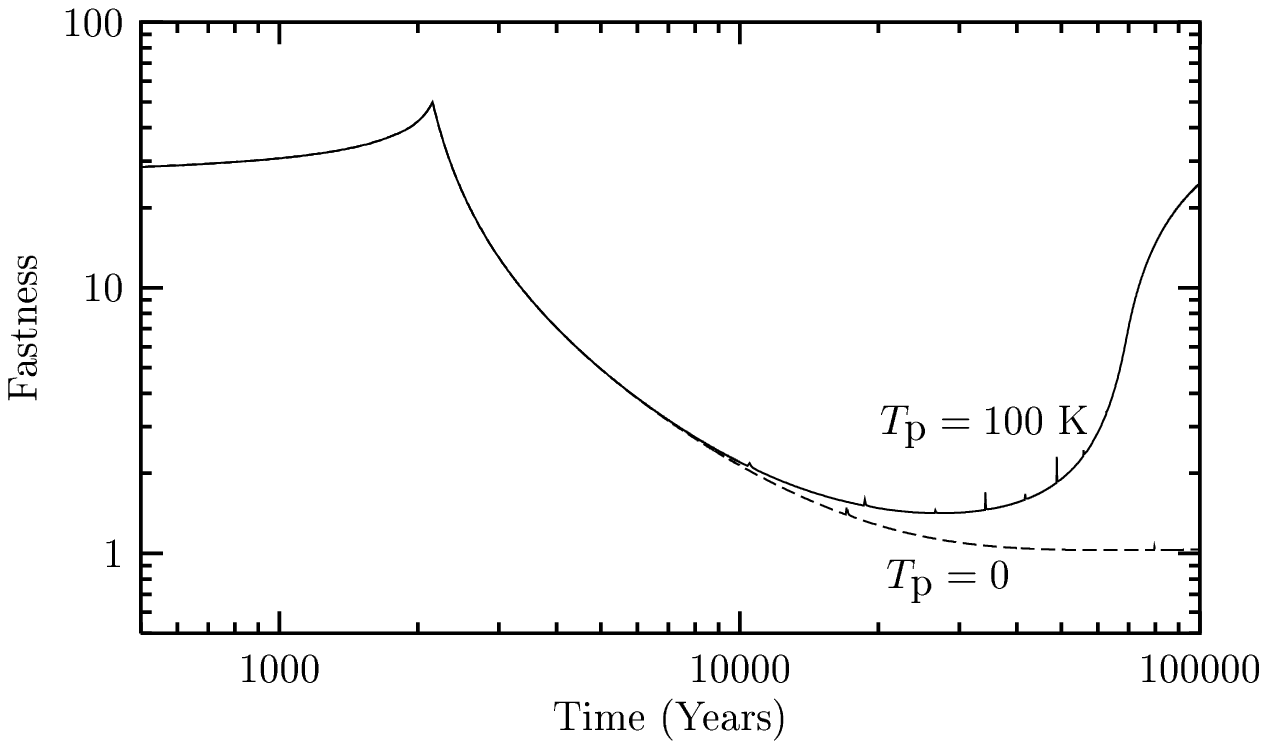}{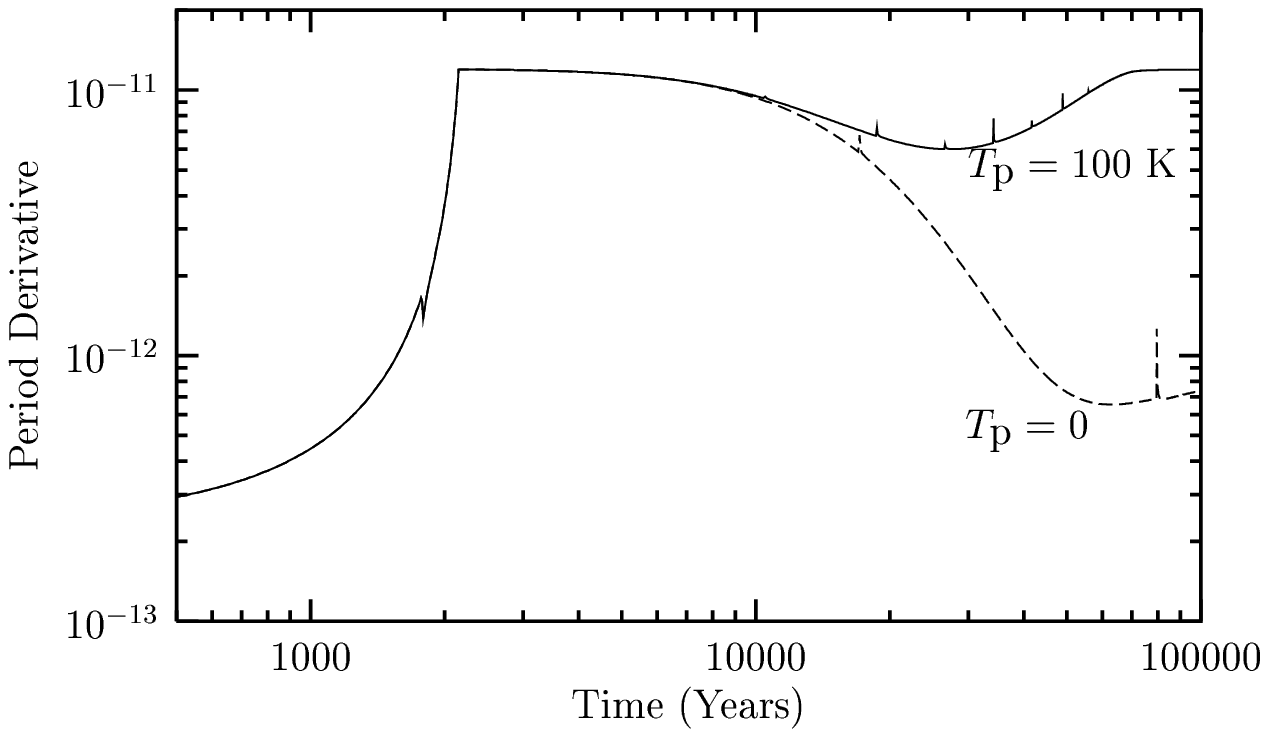}
%\vspace{-12.5 cm}
\caption{(Left panel) The period-derivative evolution with  
$\Tp=0$ and $\Tp= 100$ K. For both models, 
$M_0 = 1 \times 10^{-4} \Msun$ and $P_0=30$ ms. (Right panel) Evolution of the fastness parameter for the same model as given in the left panel}
\label{fig10}
\end{figure}

By means of a series of numerical simulations of long-term evolution of the
neutron stars interacting with fallback disks we have shown that: 
(1) The torque model  $N_\ast \propto (1 - \omega_{\ast}^{2})$~ leads to the observed period clustering for a wide range of initial conditions provided the neutron star gets into and remains in an epoch of accretion with spin-down, which holds when the inner edge of the accretion disk penetrates the light cylinder. 
(2) The correct identification of $t_0$  yields long initial disk evolutionary timescale $t_0 \sim$ a few $10^3$ y, without requiring excessive values of $\Mdot$ and the disk mass. A wide range of initial conditions do lead to the epoch of accretion with spin-down. 
(3) The introduction of $\Tp$, the physically expected criterion for the disk to become neutral and passive starting from its outer regions, leads to evolutionary models remaining in the AXP/SGR X-ray luminosity range at the right ages when the period evolution has converged into the observed range of period clustering.
We find the minimum temperature $\Tp$ for an active disk to be $\Tp \sim 100$ K. This value is in agreement with the results of Inutsuka and Sano (2005), who concluded that turbulence works down to temperatures as low as $\sim$300 K.
The low values of $\Tp$ indicates that the inner fallback disks are active up to a large radius greater than about $10^{12}$ cm.  
(4) The period derivatives are also concur with observed values.       
(5) For dipole field strength $\ga ~10^{13}$ G, we see that the model sources cannot evolve into the AXP phase consistently with observed luminosity, period and period-derivatives. Based on this result, we expect that sources with dipole fields of  $\ga ~10^{13}$ G are not likely to become AXPs, even if they are born with fallback disks.
(6) The transient AXPs and SGRs with X-ray luminosities $\Lx \sim 10^{33}$ erg s$^{-1}$ in quiescence are likely to be older than the persistent sources. This is because fallback disks that have accretion rates corresponding to these luminosities are unlikely to penetrate the light cylinders of young neutron stars with  small initial spin periods. The number of observable transient AXP/SGRs are expected to be comparable to those of persistent sources, as the models indicate that the transient AXP/SGRs are in the cutoff phase of the X-ray luminosities for a wide range of initial conditions (see Figure~\ref{fig8}).

These results could provide an opportunity to study the evolutionary connection between AXP/SGRs and other young neutron star populations that could  also be evolving with active fallback disks, as proposed originally by \citet{alpar01}. This will be the subject of our future work, in particular on comparisons with the central compact objects (CCOs) and the dim isolated neutron stars (DINs).

\acknowledgements

We acknowledge research support T\"{U}B{\.I}TAK (The Scientific and Technical Research Council of Turkey) through grant 107T013. \"{U}.E and M.A.A. acknowledge support from the Sabanc\i\ University Astrophysics and Space Forum. M.A.A. thanks The Turkish Academy of Sciences for research support. This work has been supported by the Marie Curie EC FPG Marie Curie Transfer of Knowledge Project ASTRONS, MKTD-CT-2006-042722. We thank Graham Wynn for usefull discussions and comments on the manuscript.

\clearpage

\begin{deluxetable}{ccrrrrrrrrcrl}
\tabletypesize{\scriptsize}
\tablecaption{Values of some constants in two different opacity regimes \label{table1}}
\tablewidth{0pt}
\tablehead{
\colhead{Opacity} & \colhead{$C$} & \colhead{$p$} & \colhead{$q$} & \colhead{$\alpha$} &
\colhead{$\gamma_1$} & \colhead{$\gamma_2$}
}
\startdata
$\kappa_{\rm es}\gg \kappa _{\rm bf}$ & $\left(\frac{27\kappa_0}{GM\sigma}%
\right)^{1/3}\left(\frac{\alpha_{\rm ss}k_{\rm B}}{\mu m_{\rm p}}\right)^{4/3}$ & 1 &  
2/3 & 19/16 & $3.6851 \times 10^{-4}$ & $6.2626 \times 10^{-4}$ \\ 
$\kappa_{\rm bf}\gg \kappa _{\rm es}$ & $\alpha _{\rm ss}^{\frac{8}{7}}\left(\frac{%
27\kappa_0}{\sigma}\right) ^{\frac{1}{7}}\left(\frac{k_{\rm B}}{\mu m_{\rm p}}%
\right)^{\frac{15}{14}}\left(GM\right)^{-\frac{5}{14}}$ & 15/14 & 3/7 & 5/4 & $1.7030 \times 10^{-5}$ & $3.3558 \times 10^{-5}$ \\
\enddata

%% Text for table notes should follow after the \enddata but before
%% the \end{deluxetable}. Make sure there is at least one \tablenotemark
%% in the table for each \tablenotetext.

\tablecomments{The values of the constants in two different opacity regimes. Here, 
$\kappa_{es}$ is the electron scattering opacity and $\kappa_{bf}$ is the bound-free opacity.
$\alpha_{ss}$ is the Shakura Sunyaev viscosity parameter, $\mu$ is the mean molecular weight,
$M$ is the mass of the neutron star which we assume to be $1.4M_{\sun}$.}

\end{deluxetable}

\clearpage

\clearpage

\clearpage

\clearpage

\clearpage

\clearpage

\clearpage

\clearpage

\clearpage

\clearpage

\clearpage

\clearpage

\end{document}